\documentclass[aps,twocolumn]{revtex4}

\RequirePackage[T2A]{fontenc}

\usepackage{amssymb}
\usepackage{amsmath}
\usepackage{bm}
\usepackage[dvips]{epsfig}
\usepackage{graphicx}
\usepackage{textcomp}

\begin{document}

\title{On the types of quasiperiodic potentials and level surfaces 
of quasiperiodic functions} 

\author{A.Ya. Maltsev}

\affiliation{
\centerline{\it L.D. Landau Institute for Theoretical Physics}
\centerline{\it 142432 Chernogolovka, pr. Ak. Semenova 1A}}

\begin{abstract}
We examine the similarities and differences between two types 
of quasiperiodic functions: those obtained by restricting smooth 
periodic functions in $\mathbb{R}^N$ to lower-dimensional subspaces 
$\mathbb{R}^n \subset \mathbb{R}^N$, and those that arise as physical 
fields or potentials in quasicrystalline structures. Our analysis 
reveals that functions in these two classes share a number of common 
features while also exhibiting significant differences. As a key tool 
for investigating both classes of functions, we employ the Novikov 
problem - specifically, the problem of describing the global geometry 
of the level sets of quasiperiodic functions.
\end{abstract}

\maketitle

\section{Introduction}

 This work is devoted to the study of functions belonging to certain 
types of quasiperiodic functions in $\mathbb{R}^{n}$. It should be noted 
that, in the most general sense, there exists a variety of distinct
classes of quasiperiodic functions (see, for example, \cite{Bohr1,Bohr2,Stepanoff1,Stepanoff2,Weyl,Besicovitch1,Besicovitch2}), 
each possessing its own specific characteristics.

 Here, we shall call a function $f({\bf x}) = f(x^1, \dots, x^n)$ a 
quasiperiodic function with $N$ quasiperiods if it is the restriction 
of an $N$-periodic function $F({\bf X}) = F(X^1, \dots, X^N)$ to 
$\mathbb{R}^n$ for some affine embedding
$\, \mathbb{R}^{n} \subset \mathbb{R}^{N} \, $ :
$$f (x^{1}, \dots , x^{n} ) \,\,\, = \,\,\, 
F \left( X^{1} , \dots , X^{N} \right) 
\big|_{\mathbb{R}^{n} \subset \mathbb{R}^{N}} $$

 We are interested in describing the level surfaces of functions 
$f(\mathbf{x})$ (a generalized version of Novikov's problem). 
We therefore assume that the functions $f(\mathbf{x})$ are smooth 
and may possess only specific types of singularities - namely, 
multiple saddle points or isolated maxima and minima. Level 
surfaces $f(\mathbf{x}) = \text{const}$ that contain no singular 
points of $f(\mathbf{x})$ are referred to as non-singular (or regular) 
level surfaces, while those that contain singular points of 
$f(\mathbf{x})$ are called singular level surfaces.

\vspace{1mm}

  The most regular class of quasiperiodic functions on $\mathbb{R}^n$ 
with $N$ quasiperiods consists of the restrictions of smooth functions 
$F(X^1, \dots, X^N)$ to an affine subspace 
$\mathbb{R}^n \subset \mathbb{R}^N$ with an irrational orientation. 
Here, we shall refer to such functions as functions of type A.

 Quasiperiodic functions of type A play an important role in a multitude 
of crucial problems in theoretical and mathematical physics, as well as 
in a wide range of applications. One of the most fundamental problems 
associated with the theory of such functions is that of describing open 
(non-closed) level lines
\begin{equation}
\label{LevelLines}
f (x, y) \,\,\, = \,\,\, c 
\end{equation}
of quasiperiodic functions of type A in the plane (the Novikov problem).

 The most important case here is $N = 3$, which is related to the theory 
of galvanomagnetic phenomena in strong magnetic fields. In this case,  
the functions $f(x, y)$ are the restrictions of a (3-periodic) dispersion 
relation $\epsilon(p_1, p_2, p_3)$ to planes orthogonal to the magnetic field; 
so, they are quasiperiodic functions of type A on $\mathbb{R}^2$ with three 
quasiperiods. The problem of describing the global geometry of the open level 
lines of such functions was posed by S.P. Novikov in 
\cite{MultValAnMorseTheory} and subsequently investigated extensively 
in his topological school (see 
\cite{zorich1,dynn1992,Tsarev,dynn1,zorich2,DynnBuDA,dynn2,dynn3,
PismaZhETF,ZhETF1997,UFN,BullBrazMathSoc,StatPhys,DeLeoPhysLettA,DeLeoPhysB,
TrMIAN,DeLeo2017,DynMalNovUMN}). This case is particularly interesting because 
the main type of open level lines (\ref{LevelLines}) encountered here is given 
by the ``topologically regular'' level lines.

\begin{figure}[t]
\begin{center}
\includegraphics[width=\linewidth]{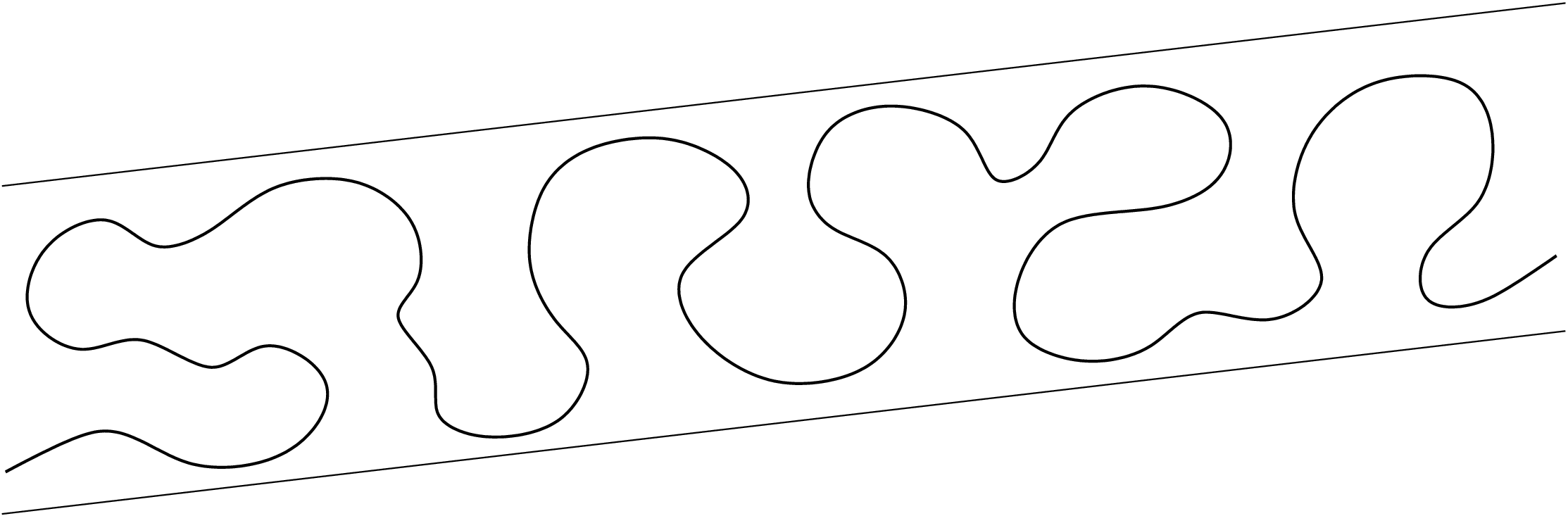}
\end{center}
\caption{A topologically regular level line (\ref{LevelLines}) 
(schematic representation).
}
\label{TopRegular}
\end{figure}

 Topologically regular level lines possess remarkable geometric properties. 
Specifically, every topologically regular open level line (\ref{LevelLines}) 
lies within a straight strip of finite width, passing through it 
(Fig. \ref{TopRegular}).

 In the case $N = 3$, only the topologically regular level lines 
(\ref{LevelLines}) are open level lines that remain stable under small 
variations of the problem's parameters (\cite{zorich1,dynn1}). 
Furthermore, each stable family of such level lines corresponds to 
specific topological invariants; this served as the basis for 
introducing the topological numbers observed in the conductivity 
of normal metals in \cite{PismaZhETF}.

\begin{figure}[t]
\begin{center}
\includegraphics[width=\linewidth]{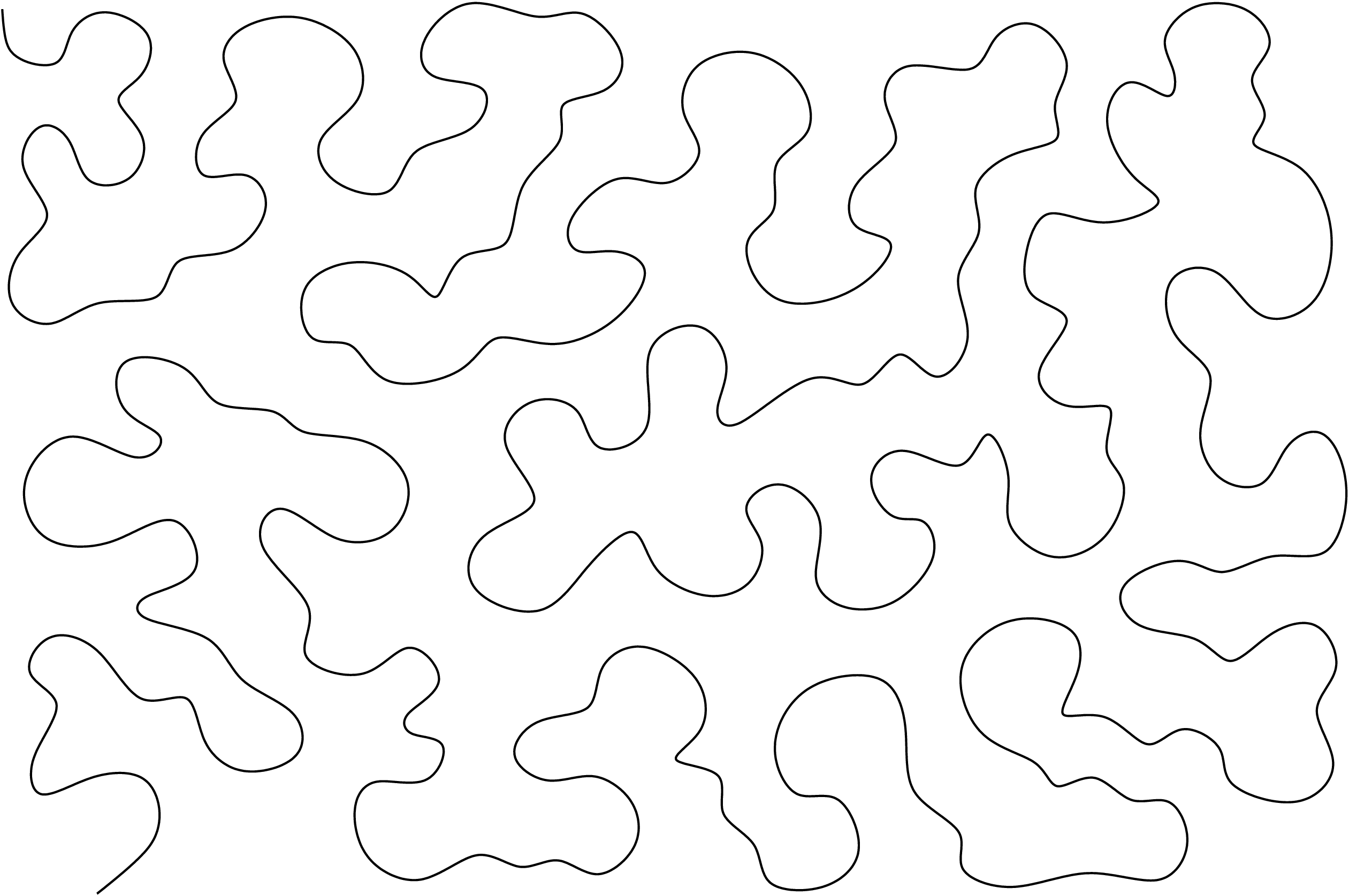}
\end{center}
\caption{Dynnikov’s chaotic level line (schematic).}
\label{Chaotic}
\end{figure}

 In addition to topologically regular open level lines (\ref{LevelLines}), 
``chaotic'' level lines (of the Tsarev or Dynnikov type) may also appear; 
these have a more complex shape (Fig. \ref{Chaotic}). In the case $N = 3$, 
however, their emergence imposes non-trivial conditions on the function 
$f(x, y)$ (\cite{dynn3}).

 More precisely, let us consider a smooth 3-periodic function 
$F(X, Y, Z)$ and fix the direction of the magnetic field $\mathbf{B}$.
Consider a plane $\Pi(\mathbf{B})$ that is orthogonal to $\mathbf{B}$, 
and the level lines (\ref{LevelLines}) for the function
$$f (x, y) \,\,\, = \,\,\, F \left( X, Y, Z \right)
\big|_{\Pi ({\bf B}) \subset \mathbb{R}^{3}} $$

 Let us assume that the direction $\, \Pi ({\bf B}) \, $ is not 
rational - that is, it is not generated by two periods of 
$\, F \left( X, Y, Z \right) \, $ in $\, \mathbb{R}^{3} \, $. 
Then, according to \cite{dynn3}:

\vspace{2mm}

\noindent
1) Open level lines (\ref{LevelLines}) arise (in all planes 
$\Pi ({\bf B}) \perp {\bf B}$) for all values of $c$ within a certain
connected interval
$$c \in \left[ c_{1} ({\bf B}) , \, c_{2} ({\bf B}) \right] $$
($c_{2} ({\bf B}) \geq c_{1} ({\bf B})$). 

\vspace{2mm}

\noindent
2) Chaotic level lines (\ref{LevelLines}) can arise only in the case 
$ \, c_{1} ({\bf B}) = c_{2} ({\bf B}) \, $.

\vspace{2mm}

 It should be said that the study of chaotic level lines (\ref{LevelLines}) 
for $N = 3$ is of great interest and continues at the present time 
(see, for example,
\cite{zorich3,DeLeo1,DeLeo2,DeLeo3,DeLeoDynnikov1,Dynnikov2008,DeLeoDynnikov2,
Skripchenko1,Skripchenko2,DynnSkrip1,DynnSkrip2,AvilaHubSkrip1,AvilaHubSkrip2,
DynHubSkrip}). 

\vspace{2mm}

 The open level lines (\ref{LevelLines}) for the case $N = 4$ were 
investigated in the works \cite{NovKvazFunc,DynNov,AnnPhys,Superpos}. 
In general, however, the case $N = 4$ has been studied in much less 
detail compared to the case $N = 3$.

\vspace{2mm}

 When considering the general case (with arbitrary $n$ and $N$) for 
potentials of type A, it is also convenient to examine the level 
surfaces of the functions 
\begin{equation}
\label{fFConnection}
f ({\bf x}) \,\,\, = \,\,\, F \left( {\bf X} \right) 
\big|_{\mathbb{R}^{n} \subset \mathbb{R}^{N}} 
\end{equation}
simultaneously for all (affine) subspaces 
$\, \mathbb{R}^{n} \subset \mathbb{R}^{N} \, $ with a given direction 
$\, \xi \in G_{N,n} \, $. In the context of functions with $N$ quasiperiods, 
it is natural to assume that the direction $\xi$ does not lie in any 
rational hyperplane $\mathbb{R}^{N-1} \subset \mathbb{R}^{N}$. 
In this case, the following statements hold for the functions 
$\, f({\bf x}) \, $ (\cite{DynMalNovUMN,BigQuas}).

\vspace{2mm}

\noindent
1) For a fixed function $F$ and a fixed direction $\xi$, open level 
surfaces
\begin{equation}
\label{TypeALevSurf}
f ({\bf x}) \,\,\, = \,\,\, c 
\end{equation}
emerge (in at least one of the subspaces 
$\mathbb{R}^{n} \subset \mathbb{R}^{N}$ of the direction $\xi$) 
for values $c$ within a certain connected interval
$$c \,\,\, \in \,\,\, \left[ c_{1} ( \xi ) , \, c_{2} ( \xi ) \right] 
\,\,\, , $$
which can contract into a single point ($c_{1} ( \xi ) = c_{2} ( \xi )$).

\vspace{2mm}

\noindent
2) For $c \notin [c_{1}(\xi), c_{2}(\xi)]$, the sizes of all closed level 
surfaces \eqref{TypeALevSurf} - within all subspaces 
$\mathbb{R}^{n} \subset \mathbb{R}^{N}$ of direction $\xi$ - are bounded 
by a single constant $D(\xi, c)$, which may tend to infinity as $\, c \, $  
approaches the interval $\, [c_{1}(\xi), c_{2}(\xi)] \, $.

\vspace{2mm}

 As was also shown in \cite{BigQuas}, for $c \in (c_{1}(\xi), c_{2}(\xi))$, 
the open level surfaces (\ref{TypeALevSurf}) arise in all subspaces 
$\mathbb{R}^{n} \subset \mathbb{R}^{N}$ of the direction $\xi$. However, 
for $c = c_{1}(\xi)$ or $c = c_{2}(\xi)$ (or $c = c_{1}(\xi) = c_{2}(\xi)$), 
such open level surfaces may arise only in a subset of the subspaces 
$\mathbb{R}^{n} \subset \mathbb{R}^{N}$ (of direction $\xi$). 
In the latter case, these subspaces form a dense set within the 
collection of all subspaces $\mathbb{R}^{n} \subset \mathbb{R}^{N}$ 
corresponding to the direction $\xi$.

\vspace{2mm}

 In addition to the theory of galvanomagnetic phenomena in metals, 
the (generalized) Novikov problem for A-type potentials plays an important 
role in the theory of dynamical systems 
(see, e.g., \cite{SagdeevJETPLett1986,SagdeevUFN}), the theory 
of ``superpotentials'' in electronic systems 
(see, e.g., \cite{JMathPhys2004}), systems of ultracold atoms 
(\cite{DynMalAtoms}), the physics of multilayer systems 
(\cite{AnnPhys,Superpos,TitovKatsnelson}), as well as in a number of 
other scientific fields.

\vspace{2mm}
 
 In this paper, however, we wish to consider quasiperiodic functions 
$f(\mathbf{x})$ of a somewhat different type. Our primary focus will be 
the study of the open level surfaces of such functions (the generalized 
Novikov problem); therefore, we shall continue to require that the 
functions $f(\mathbf{x})$ possess a sufficient degree of smoothness. 
As before, we assume that all singular points of the functions 
$f(\mathbf{x})$ are either multiple saddle points or isolated maxima 
and minima.

 To satisfy the above requirements, however, it is not necessary 
to demand that the functions $F \left( \mathbf{X} \right)$ themselves 
be smooth; one may instead require only that the restrictions of the 
functions $F \left( \mathbf{X} \right)$ to the subspaces 
$\, \mathbb{R}^{n} \subset \mathbb{R}^{N} \, $ 
(as given by \eqref{fFConnection}) be smooth.

 Despite their seeming artificiality, functions $f(\mathbf{x})$ of this 
type arise naturally in the physics of quasicrystals (see. e.g.
\cite{KaluginKitaevLevitov1} and the references therein). The corresponding 
functions $\, F \left(\mathbf{X} \right) \, $ also emerge naturally when 
the so-called ``section method'' (\cite{OgDunKatz,Arnold1988}) is applied 
to the study of quasicrystalline structures.

 Namely, as is well known - dating back to the work of De Bruijn 
\cite{DeBruijn} - quasicrystalline structures in $\mathbb{R}^{n}$ are 
traditionally viewed as the result of projecting periodic structures 
from $\mathbb{R}^{N}$ onto a subspace $\mathbb{R}^{n}$, for some embedding 
$\mathbb{R}^{n} \rightarrow \mathbb{R}^{N}$ of an irrational direction 
$\xi \in G_{N,n}$. In particular, quasiperiodic tilings (or partitions) 
of $\mathbb{R}^{n}$ by polytopes are obtained by projecting a (connected) 
set 
$$G^{n} \,\,\, = \,\,\, \bigcup_{j} \, G^{n}_{j} $$ 
of $n$-dimensional faces of the integer lattice 
$\mathbb{Z}^{N} \subset \mathbb{R}^{N}$ onto the subspace $\mathbb{R}^{n}$.

\vspace{1mm}

\begin{figure}[t]
\begin{center}
\includegraphics[width=\linewidth]{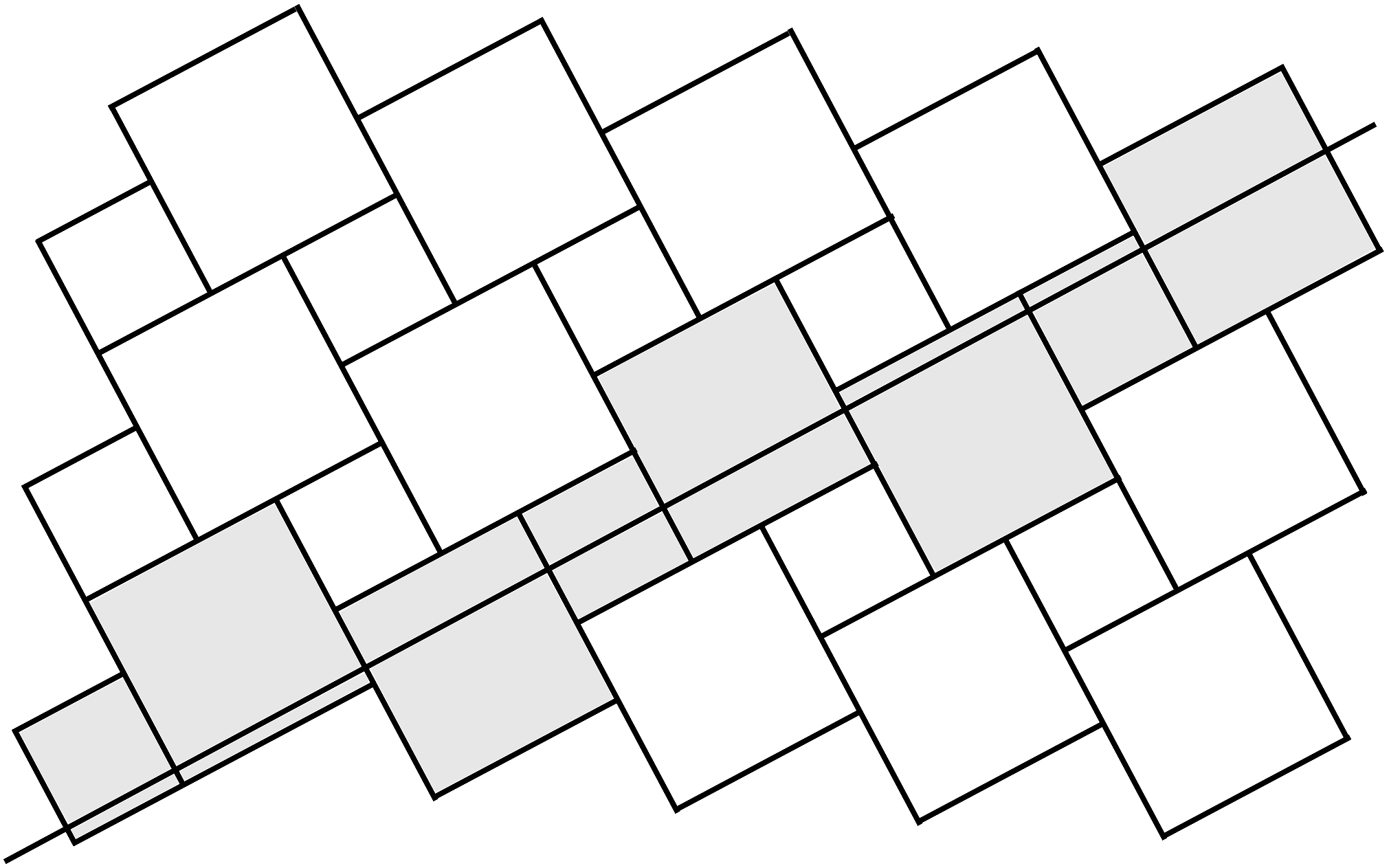}
\end{center}
\caption{A periodic tiling of the space $\mathbb{R}^{N}$ by polytopes 
$\widehat{M}^{N}_{i,j}$, which defines a quasicrystalline structure 
in $\mathbb{R}^{n} \subset \mathbb{R}^{N}$ (schematic representation).}
\label{Razb}
\end{figure}

 The section method allows one to define the same partition as the result 
of the intersection of $\, \mathbb{R}^{n} \, $ with the polytopes 
$\, \widehat{M}^{N}_{i,j} \subset \mathbb{R}^{N} \, $, which form a periodic 
partition of the space $\, \mathbb{R}^{N} \, $ (Fig. \ref{Razb}).

\vspace{1mm}

 This partition corresponds to a partition of the torus
\begin{equation}
\label{MNi}
\mathbb{T}^{N} \,\,\, = \,\,\, 
\mathbb{R}^{N} \big/ \, \mathbb{Z}^{N} 
\end{equation}
into a finite number of polytopes
$\, \left\{ M^{N}_{i} , \,\,\, i = 1, \dots , I \right\} $:
\begin{equation}
\label{RazbienieTora}
\mathbb{T}^{N} \,\,\, = \,\,\, \bigcup_{i} \, M^{N}_{i} \,\,\, , 
\end{equation}
each of which is of the form
$$M^{N}_{i} \,\,\, = \,\,\, P^{n}_{i} \, \times \, Q^{N-n}_{i} \,\,\, , $$
where the polytope $\, P^{n}_{i} \, $ lies in an $n$ - dimensional plane
along the direction $\, \xi \, $, and the polytope $\, Q^{N-n}_{i} \, $
lies in an $(N-n)$ - dimensional plane along the direction $\, \xi^{\perp} \, $.

 The interiors of the polytopes $M^{N}_{i}$ are disjoint, while distinct 
$M^{N}_{i}$ may touch one another along faces (of various dimensions).

\vspace{1mm}

 As we have already said, the partition (\ref{RazbienieTora}) corresponds 
to a periodic partition of the space $\mathbb{R}^{N}$ into polytopes
\begin{equation}
\label{RazbienieRN}
\mathbb{R}^{N} \,\,\,  = \,\,\, \bigcup_{i,j} \, 
{\widehat{M}}^{N}_{i,j} \,\,\, = \,\,\, \bigcup_{i,j} \, 
{\widehat{P}}^{n}_{i,j} \, \times \, {\widehat{Q}}^{N-n}_{i,j} \,\,\, ,
\end{equation}
$(j = 1, \dots , \infty) \, , \,\, $
among which there are only a finite number of distinct ones.

\vspace{1mm}

 The periodic partition (\ref{RazbienieRN}) remains invariant under parallel 
shifts of the space $\, \mathbb{R}^{n} \, $ within $\,\mathbb{R}^{N} \, $. 
However, both the set of polytopes that intersect $\, \mathbb{R}^{n} \, $ and 
the set of faces $\, G^{n}_{j} \, $ belonging to $ \, G^{n} \, $ do change.

 Changes in a quasicrystalline structure in $\mathbb{R}^{n}$ 
occur ``instantaneously''; they are ``local'' in nature and involve 
the replacement of one $\, \widehat{M}^{N}_{i,j} \, $ (or a finite set of 
polytopes $\, \widehat{M}^{N}_{i,j}$) with another polytope 
$\, \widehat{M}^{N}_{i^{\prime},j^{\prime}} \, $ (or a finite set of such 
polytopes $\, \widehat{M}^{N}_{i^{\prime},j^{\prime}}$) in the intersection 
with $\mathbb{R}^{n}$. In the physics of quasicrystals, such changes 
are referred to as phase transformations (or phasons). It should also be noted 
that, at the very moment of such a change, the specific form of the quasicrystalline 
structure (and the corresponding subset of $\, \widehat{M}^{N}_{i,j}$) is determined 
by an additional rule within the chosen model.

\vspace{1mm}

 Here, we will base our approach specifically on the partition of the 
torus (\ref{RazbienieTora}) and the corresponding periodic partition of 
the space $\mathbb{R}^{N}$ (\ref{RazbienieRN}).

\vspace{1mm}

 As is easy to see, all $(N-1)$-dimensional faces of the polytopes 
$\, M^{N}_{i} \, $ lie either in a plane containing the direction 
$\, \xi \, $ or in a plane containing the direction $\, \xi^{\perp} \, $. 
We shall say that faces of the first type are parallel to the direction 
$\, \xi \, $, while faces of the second type are orthogonal to the 
direction $\, \xi \, $.

\vspace{1mm}

 In the general case, the faces of $\, M^{N}_{i} \, $ (of dimension less 
than $N-1$) may be neither parallel nor orthogonal to $\, \xi \, $. 

\vspace{1mm}

 As usual, we consider a face of $\, M^{N}_{i} \, $ to be parallel 
to $\, \xi \, $ if its direction either contains $\, \xi \, $ 
(if its dimension is $\, \geq n$) or is contained within $\, \xi \, $ 
(if its dimension is $\, \leq n$). Faces of the first type have the form
$$P^{n}_{i} \, \times \, q^{s}_{i} \,\,\, , $$
where $\, q^{s}_{i} \, $ is any face of $\, Q^{N-n}_{i} \, $, while 
faces of the second type are of the form
$$p^{s}_{i} \, \times \, q^{0}_{i} \,\,\, , $$
where $\, p^{s}_{i} \, $ is any face of $\, P^{n}_{i} \, $ 
(or $\, P^{n}_{i} \, $ itself) and $\, q^{0}_{i} \, $ is any vertex 
of the polytope $ \, Q^{N-n}_{i} \, $.

\vspace{1mm}

 Similarly, a face of $\, M^{N}_{i} \, $ is orthogonal to $\, \xi \, $ 
if its direction either contains $\, \xi^{\perp} \, $ 
(its dimension is $\, \geq (N-n)$) or is contained within $\, \xi^{\perp} \, $ 
(its dimension is $\, \leq (N-n)$). Faces of the first type have the form
$$p^{s}_{i} \, \times \, Q^{N-n}_{i} \,\,\, , $$
where $\, p^{s}_{i} \, $ is any face of $\, P^{n}_{i} \, $, 
while faces of the second type are of the form
$$p^{0}_{i} \, \times \, q^{s}_{i} \,\,\, , $$
where $\, p^{0}_{i} \, $ is any vertex of the polytope $\, P^{n}_{i} \, $, 
and $\, q^{s}_{i} \, $ is any face of the polytope $\, Q^{N-n}_{i} \, $ 
(or $Q^{N-n}_{i}$ itself).

\vspace{1mm}

 In the general case, each face of a polytope $\, M^{N}_{i} \,$ belongs 
either to an $(N-1)$ - dimensional face (hyperface) of $\, M^{N}_{i}$ that 
is parallel to $\, \xi \, $, or to an $(N-1)$ - dimensional face that is 
orthogonal to $\, \xi \, $ (or to both).

\vspace{1mm}

 As is not difficult to see, changes in the quasicrystalline structure 
in $\, \mathbb{R}^{n} \subset \mathbb{R}^{N} \, $ occur at the moments when 
$\mathbb{R}^{n}$ intersects the faces of $\, \widehat{M}^{N}_{i,j} \, $ that 
are parallel to the direction $\, \xi \, $.

\vspace{2mm}

 The quasiperiodic properties of quasicrystals are typically described 
using functions $\, f(\mathbf{x}) \, $ associated with a quasicrystalline 
tiling (or covering) of $\mathbb{R}^{n}$. Such functions are generally 
constructed from a finite set of functions $\, f_{i}(\mathbf{x}) \, $, 
each defined on a corresponding tile (or element of the covering) $\, O_i \, $. 
These functions may, for instance, be constant on each $\, O_i \, $ 
(\cite{Arnold1988}), vanish at the boundaries of $\, O_i \, $ 
(see, e.g., \cite{NovFom}), or possess other specific properties. The
functions $\, f(\mathbf{x}) \, $ can be also defined as the restrictions 
\eqref{fFConnection} of certain functions $\, F(\mathbf{X}) \, $ that 
satisfy corresponding conditions on the polytopes $\, \widehat{M}^N_{i,j} \, $ 
(\cite{Arnold1988}). The spectrum of the functions $\, f(\mathbf{x}) \, $ 
defines a lattice of Fourier harmonics, 
$\, \mathbb{Z}^{N}_{*} \subset \mathbb{R}^{n} \, $, which is directly 
observable in X-ray scattering experiments involving quasicrystals. 
It is easy to see that this lattice 
$\, \mathbb{Z}^{N}_{*} \subset \mathbb{R}^{n}$ remains the same for all 
embeddings $\, \mathbb{R}^{n} \subset \mathbb{R}^{N} \, $ corresponding 
to a given direction $\, \xi \in G_{N,n} \, $.

 Let us note once again that, at the very moment of a change of the 
quasicrystalline structure, the function $\, f({\bf x}) \, $ is defined 
by the intersection with a polytope $\, {\widehat{M}}^{N}_{i,j} \, $ 
(or a set $\{{\widehat{M}}^{N}_{i,j}\}$) - or, alternatively, 
with $\, {\widehat{M}}^{N}_{i^{\prime},j^{\prime}} \, $ 
(or $\, \{{\widehat{M}}^{N}_{i^{\prime},j^{\prime}}\}$) - in accordance 
with a specific supplementary definition corresponding to the chosen model. 
Certainly, this supplementary definition is of purely theoretical 
significance and has no effect on the position of the 
lattice $\, \mathbb{Z}^{N}_{*} \, $. 

\vspace{2mm}

 The functions $f(\mathbf{x})$ that arise in the physics of quasicrystals 
are smooth in the space $\mathbb{R}^{n}$. At the same time, they are 
determined by the quasicrystalline structure in $ \, \mathbb{R}^{n} \, $ 
and undergo abrupt changes during phase transformations.

\vspace{1mm}

\begin{figure}[t]
\begin{center}
\includegraphics[width=\linewidth]{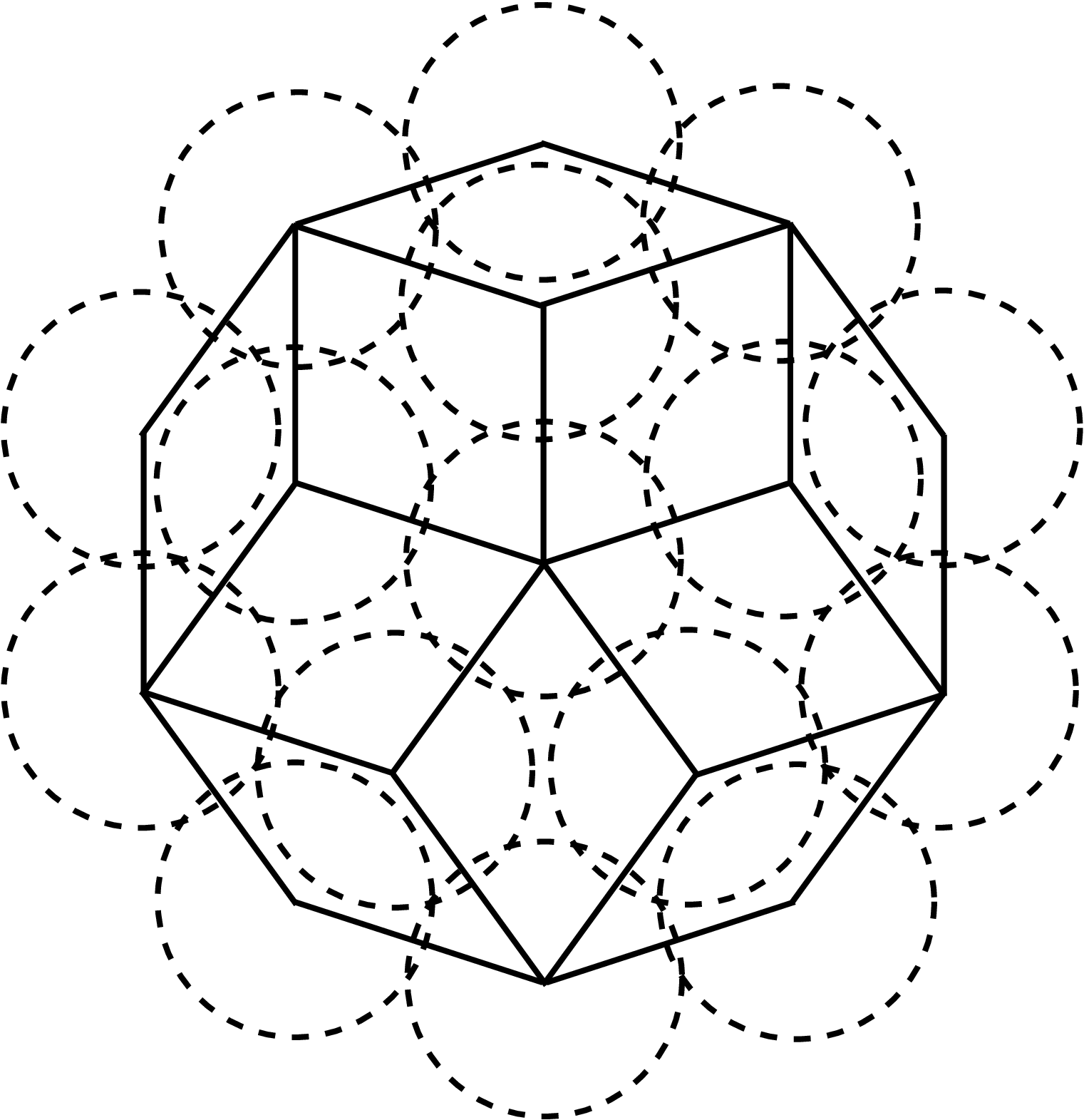}
\end{center}
\caption{A smooth potential generated by short-range potentials of 
atoms in a quasicrystalline structure (schematic representation).}
\label{QuasCrystPot}
\end{figure}

 As a simple example, consider the potential $V(\mathbf{x})$ 
generated by atoms that form a quasicrystalline structure. Let us have 
a quasicrystalline tiling, defined by an embedding 
$\, \mathbb{R}^n \to \mathbb{R}^N \, $ and the associated partitions 
\eqref{RazbienieTora}-\eqref{RazbienieRN}. Suppose that atoms are 
located at the vertices of this tiling, each generating a potential 
$V_{i}(\mathbf{x})$. We assume that these (identical) potentials, 
$V_{i}(\mathbf{x}) = V_{0}(\mathbf{x} - \mathbf{x}_{i})$, 
are smooth and have compact supports that do not extend beyond 
the polytopes adjacent to the corresponding vertex 
(Fig. \ref{QuasCrystPot}).

 The potential
\begin{equation}
\label{SummaVi}
V ({\bf x}) \,\,\, = \,\,\, \sum_{i} \, V_{i} ({\bf x}) \,\,\, , 
\end{equation}
is evidently smooth, and its values on each element of the tiling
are uniquely determined by the type of that element. However, parallel 
shifts of $\, \mathbb{R}^{n} \, $ within $\, \mathbb{R}^{N} \, $ induce 
sharp local ``jumps'' in the potential $\, V ({\bf x}) \, $, which 
correspond to local phase transformations of the quasicrystalline 
structure (phasons). It is easy to see that the potential 
$\, V ({\bf x}) \, $ can be represented as the restriction
\begin{equation}
\label{VOgran}
V ({\bf x}) \,\,\, = \,\,\, \widehat{V} \left( {\bf X} \right)  
\big|_{\mathbb{R}^{n} \subset \mathbb{R}^{N}} \,\,\, , 
\end{equation}
where $\, \widehat{V} \left( {\bf X} \right) \, $ is a function that is 
smooth within each polytope $\, {\widehat{M}}^{N}_{i,j} \, $ - as well as 
on its faces that are orthogonal to the embedding 
$\, \mathbb{R}^{n} \rightarrow \mathbb{R}^{N} \, $ - but which exhibits 
discontinuities across faces that are parallel to the embedding 
$\, \mathbb{R}^{n} \rightarrow \mathbb{R}^{N} \, $.

\vspace{1mm}

\begin{figure}[t]
\begin{center}
\includegraphics[width=\linewidth]{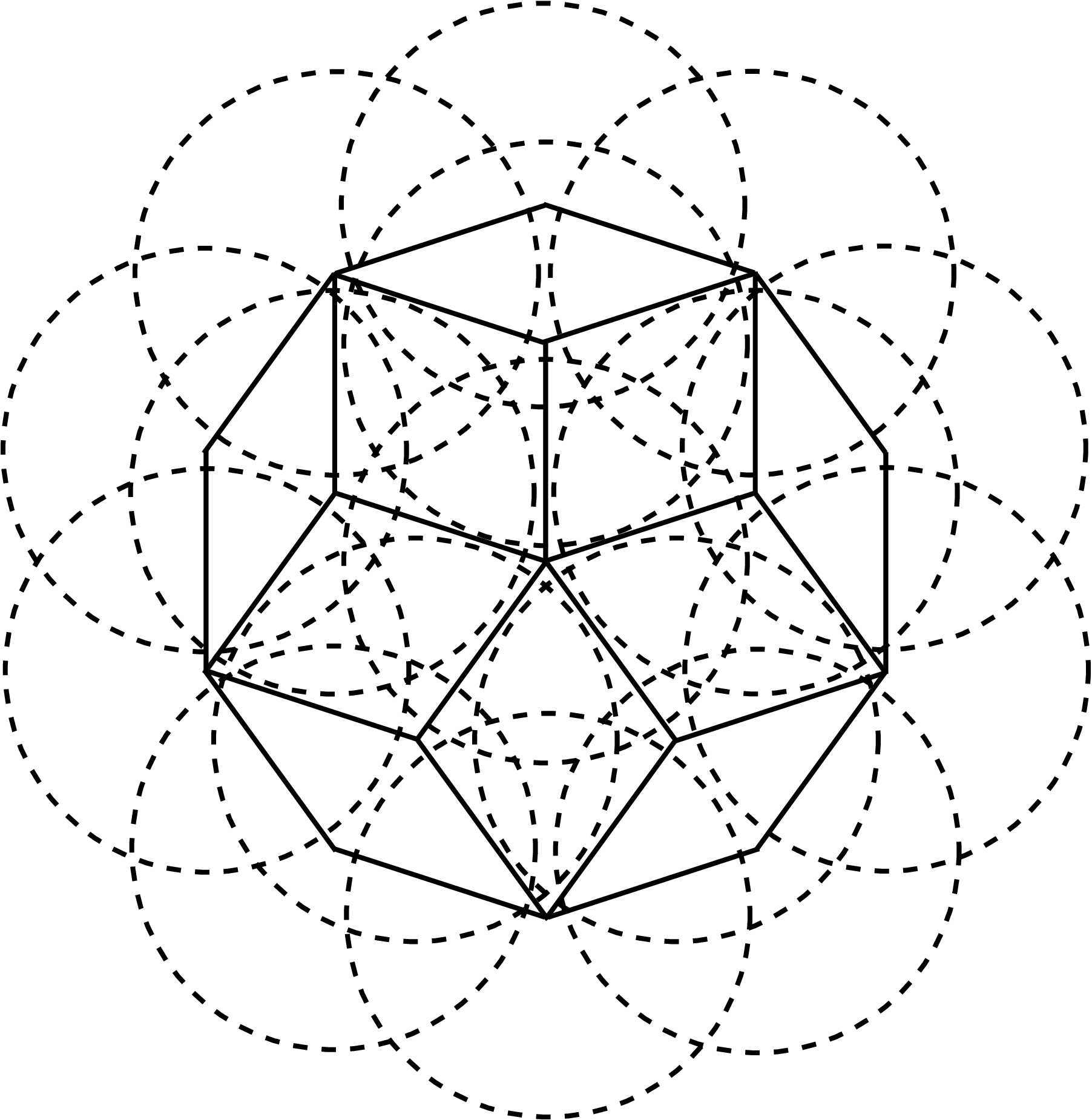}
\end{center}
\caption{A smooth potential $V(\mathbf{x})$ generated by atomic 
potentials with a long (finite) range (schematic representation).}
\label{QuasCrystLongPot}
\end{figure}

 In a more general case, one can consider smooth potentials 
$\, V_{i}(\mathbf{x}) = V_{0}(\mathbf{x} - \mathbf{x}_{i}) \, $ with 
supports of large (but finite) radius (Fig. \ref{QuasCrystLongPot}).

 In this case, the values of potential (\ref{SummaVi}) within each
of the tiling cells depend not only on the type of a cell but also on 
its surroundings.

 At the same time, it remains convenient to represent 
$V(\mathbf{x})$ in the form (\ref{VOgran}) for a modified partition 
(\ref{RazbienieTora})-(\ref{RazbienieRN}). To construct such a partition, 
let us refer back to the initial partition (\ref{RazbienieRN}).

 Each cell $\, O_{k} \, $ of the tiling
\begin{equation}
\label{RnOk}
\mathbb{R}^{n} \,\,\, = \,\,\, \bigcup_{k} \, O_{k}
\end{equation}
corresponds to the intersection of $\, \mathbb{R}^{n} \, $ with a certain 
fixed polytope
$${\widehat{M}}^{N}_{i_{k}, j_{k}} \,\,\, = \,\,\, 
{\widehat{P}}^{n}_{i_{k}, j_{k}} \, \times \, 
{\widehat{Q}}^{N-n}_{i_{k}, j_{k}} \,\,\,\,\, \subset \,\,\,\,\, 
\mathbb{R}^{N} $$

 Consider the set $\Omega_{r}$ of points in $\mathbb{R}^{n}$ whose 
distance from $\, O_{k} \, $ is at most $r$. Clearly, for a fixed tiling 
(\ref{RnOk}), there are only a finite number of ways to divide the region 
$\, \Omega_{r} \, $ (into subregions) using the cells $\, O_{k^{\prime}} \, $. 
As can be seen, the various tiling patterns of $\, \Omega_{r} \, $ correspond 
to a partition of the polytope $\, \widehat{Q}^{N-n}_{i_{k},j_{k}} \, $ 
into a finite number of ``subpolytopes'':
$${\widehat{Q}}^{N-n}_{i_{k}, j_{k}} \,\,\, = \,\,\, 
\bigcup_{l_{k}} \, {\widehat{Q}}^{N-n}_{i_{k}, j_{k}, l_{k}} \,\,\, , $$
such that each tiling pattern arises in the case of the intersection of 
$\, \mathbb{R}^{n} \, $ with the corresponding polytope
$${\widehat{M}}^{N}_{i_{k}, j_{k}, l_{k}} \,\,\, = \,\,\, 
{\widehat{P}}^{n}_{i_{k}, j_{k}} \, \times \, 
{\widehat{Q}}^{N-n}_{i_{k}, j_{k}, l_{k}} $$

\vspace{1mm}

 The partitions of the polytopes $\, {\widehat{Q}}^{N-n}_{i_{k}, j_{k}} \, $ 
into subpolytopes $\, {\widehat{Q}}^{N-n}_{i_{k}, j_{k}, l_{k}} \, $
exhibit periodicity and, consequently, define a new partition of the torus 
$\, \mathbb{T}^{N} = \mathbb{R}^{N} \big/ \mathbb{Z}^{N} \, $:
$$\mathbb{T}^{N} \,\,\, = \,\,\, \bigcup_{i, l_{i}} \, M^{N}_{i, l_{i}} 
\,\,\, = \,\,\, \bigcup_{i, l_{i}} \, P^{n}_{i} \, \times \, 
Q^{N-n}_{i, l_{i}} $$
along with a partition of the space $\, \mathbb{R}^{N} \, $
\begin{equation}
\label{NewDivision}
\mathbb{R}^{N} \,\,\, = \,\,\, \bigcup_{i, l_{i}, j} \, 
{\widehat{M}}^{N}_{i, l_{i}, j} \,\,\, = \,\,\, 
\bigcup_{i, l_{i}, j} \, {\widehat{P}}^{n}_{i,j} \, \times \, 
{\widehat{Q}}^{N-n}_{i, l_{i}, j}
\end{equation}

 Assuming that the range of the potential $\, V_{0}(\mathbf{x}) \, $ is less 
than $r$, the potential $\, V(\mathbf{x}) \, $ is given, as before, by formula 
(\ref{VOgran}) with a certain periodic function $\, \widehat{V}(\mathbf{X}) \, $. 
It can be seen that the function $\, \widehat{V}(\mathbf{X}) \, $ is smooth 
within each polytope $\, \widehat{M}^{N}_{i, l_{i}, j} \, $, possesses the 
required degree of smoothness on the faces of 
$\, \widehat{M}^{N}_{i, l_{i}, j} \, $ that are orthogonal to $\, \xi \, $, 
and exhibits discontinuities on the faces parallel to $\, \xi \, $.

\vspace{1mm}

\begin{figure}[t]
\begin{center}
\includegraphics[width=\linewidth]{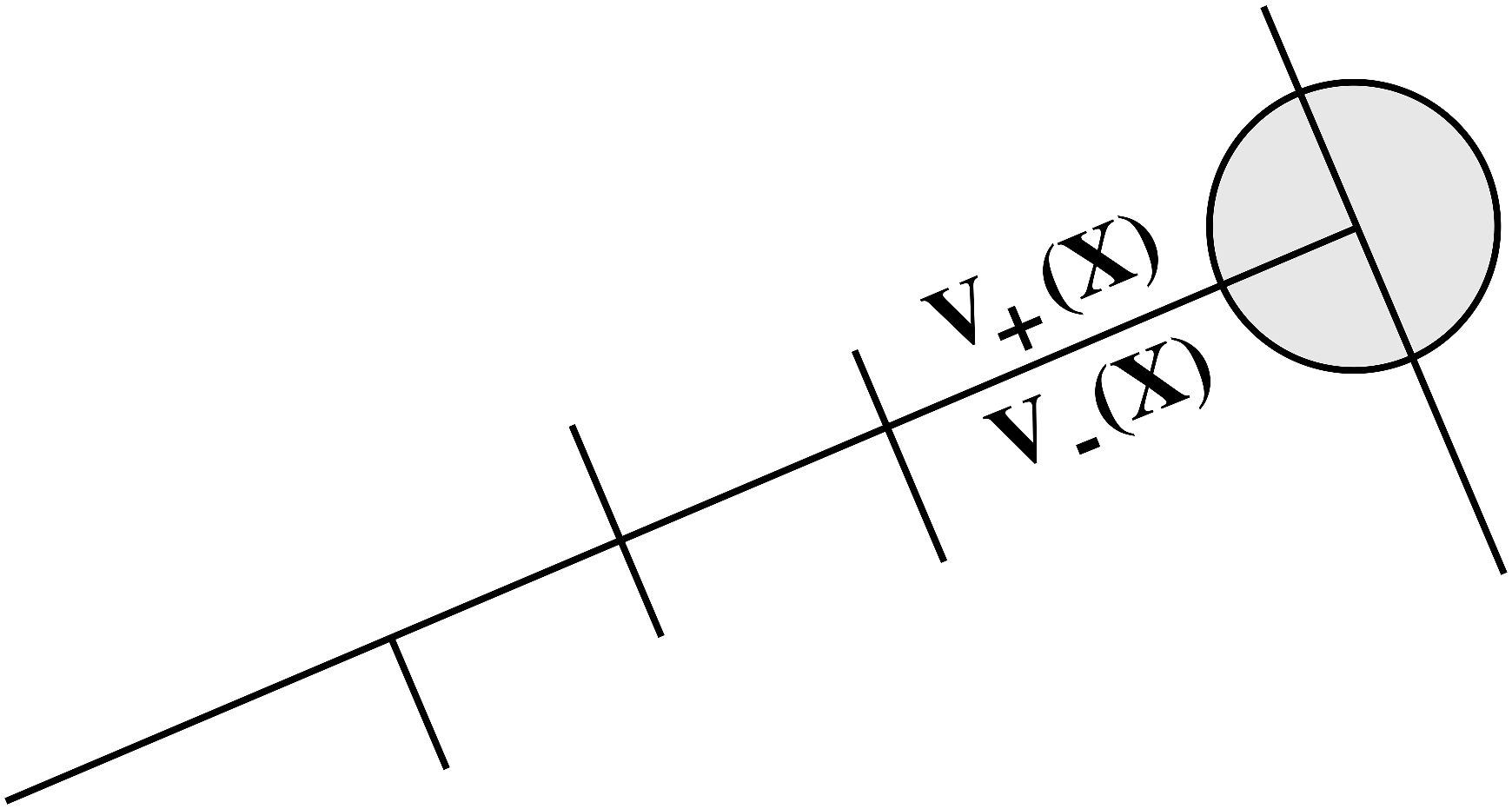}
\end{center}
\caption{The boundary of an ``extended'' ($(N-1)$-dimensional) face 
of a polytope $\, {\widehat{M}}^{N}_{i,j} \, $ and its neighborhood, 
where the function $ \, \widehat{V}(\mathbf{X}) \, $ is smooth 
(schematic representation).}
\label{ExtFaces}
\end{figure}

 In a more detailed description of the partition (\ref{NewDivision}), 
it can be noted that the intersection with any $\, (N-1)$ - dimensional 
face of $\, {\widehat{M}}^{N}_{i,j} \, $ (within the context of parallel 
shifts of $\mathbb{R}^{n}$ in $\mathbb{R}^{N}$) - where such a face is 
parallel to $\, \xi \, $ - corresponds to a local change in the 
quasicrystalline structure in $\mathbb{R}^{n}$ in the vicinity of its 
intersection with $\, {\widehat{M}}^{N}_{i,j} \, $. While this change 
has no effect on the potential $V(\mathbf{x})$ at large distances, 
it must be taken into account at distances of the order of the range 
of the potential $V_{0}(\mathbf{x})$. It can thus be seen that the new 
partition (\ref{NewDivision}) effectively corresponds to the extension 
of the faces of $\, {\widehat{M}}^N_{i,j}$ (those parallel to $\, \xi$) 
over a distance of the order of $r$, where these extensions are ``truncated'' 
at more distant $\, (N-1)$ - dimensional faces of the polytopes 
$\, {\widehat{M}}^{N}_{i^{\prime},j^{\prime}} \, $ that are orthogonal 
to $\, \xi \, $ (Fig. \ref{ExtFaces}).

 It can also be noted that, in the given construction 
(for sufficiently large $r$), the values of the function 
$\, \widehat{V}(\mathbf{X}) \, $ coincide on the upper and lower sides 
of the cut in the vicinity of the boundaries of the extended faces 
(specifically, where a face parallel to $\, \xi \, $ terminates 
at a face orthogonal to $\, \xi$; see Fig. \ref{ExtFaces}). 
Consequently, the values $\, {\widehat{V}}_{-}(\mathbf{X}) \, $ and 
$\, {\widehat{V}}_{+}(\mathbf{X}) \, $ begin to differ only at some 
distance from these boundaries. By applying the described construction - 
if necessary - we can assume this to be the case from the outset.

\vspace{2mm}

 In this work, we shall study functions $f({\bf x})$ of the 
(quasicrystalline) type described above. Like functions of type A, 
they are defined by formula (\ref{fFConnection}) and are smooth functions 
in $\mathbb{R}^{n}$. The corresponding functions $F({\bf X})$ are 
associated with the partitions (\ref{RazbienieTora})-(\ref{RazbienieRN}) 
and are assumed to be smooth everywhere except on the faces of the 
polytopes $\, {\widehat{M}}^{N}_{i,j} \, $ that are parallel to the 
direction of the embedding $\, \mathbb{R}^{n} \subset \mathbb{R}^{N} \, $. 
Furthermore, we assume that under parallel shifts of 
$\mathbb{R}^{n}$ within $\mathbb{R}^{N}$, the functions 
$f({\bf x})$ change only in accordance with changes in the 
quasicrystalline structure within $\mathbb{R}^{n}$; this imposes 
additional constraints on the functions $F({\bf X})$ 
(we shall provide precise definitions below). We shall refer to 
the corresponding functions $\, f({\bf x}) \, $ as quasiperiodic functions 
of type B or as potentials of quasicrystalline type. A primary example 
of such functions is provided by those generated by short-range atomic 
potentials in quasicrystals.

\vspace{1mm}

 It should be noted here that potentials $V(\mathbf{x})$ - 
generated by rapidly decaying atomic potentials $V_{i}(\mathbf{x})$ 
in quasicrystals - do not, strictly speaking, belong to the class 
of quasiperiodic functions of type B, even though they can be approximated 
by such functions to any desired degree of accuracy. These potentials 
can be viewed as a closure of the set of type B potentials. We shall 
refer to such functions as quasiperiodic functions of type 
$\overline{\text{B}}$, though we will not examine them in detail here. 
It is also worth noting that potentials $V(\mathbf{x})$ - 
determined by the quasicrystalline structure within a bounded 
neighborhood of a point $\mathbf{x}$ (i.e., of type B) - 
play an important role in quasicrystals that possess ``local rules'' 
(see
\cite{DeBruijn,LevitovJETP,KatzCommMathPhys,LevitovCommMathPhys,LePiunSad}). 

\vspace{2mm}

 Here, as we have already mentioned, we will be concerned with the Novikov 
problem for potentials of type B. In the next section, we will prove a number 
of statements that establish a connection between potentials of type A and 
type B in the context of the Novikov problem. In Section 3, however, we will 
provide examples that also point to significant differences between the 
quasiperiodic functions of these two types.

\section{General properties of potentials of type B from 
the perspective of the Novikov problem}
\setcounter{equation}{0}

 Here, we examine the most general properties of B-type potentials, 
defined by the formula (\ref{fFConnection}). For the sake of simplicity, 
we impose the following generic condition on the direction of the embedding
$\, \mathbb{R}^{n} \subset \mathbb{R}^{N} \, $:

\vspace{2mm}

\noindent
(g1) We assume that the direction $\, \xi \, $ intersects all integer 
subspaces $\, \mathbb{R}^{k} \subset \mathbb{R}^{N} \, $ transversally.

\vspace{2mm}

 Our reasoning will now be based on a given partition of the 
torus (\ref{RazbienieTora}) and the corresponding partition 
of $\, \mathbb{R}^{N} \, $ (\ref{RazbienieRN}).

\vspace{2mm}
 
 Let us now consider the collection $\, \left\{ M^{N}_{i} \right\} \, $ 
as an abstract set of polytopes. Let us consider each polytope
$\, M^{N}_{i} \, $ as a topological space with the topology defined by 
the natural embedding $\, M^{N}_{i} \subset \mathbb{R}^{N} \, $ 
(i.e., open sets in $\, M^{N}_{i} \, $ are defined as the intersections 
of $\, M^{N}_{i} \, $ with open sets in $\, \mathbb{R}^{N}$).

 We perform now a gluing of different
$\, M^{N}_{i} \, $ along their $(N-1)$ - dimensional faces 
(including boundaries) that are orthogonal to $\, \xi \, $, in a manner 
consistent with the gluing defined in the partition (\ref{RazbienieTora}). 
Points on the ($(N-1)$ - dimensional) faces of $\, M^{N}_{i} \, $ that are 
parallel to $\, \xi \, $ - and which do not belong to the faces orthogonal 
to $\, \xi \, $ - are to be treated as distinct (they are not identified 
with any other points). We denote the resulting complex by $\, \cal{T} \, $.

\vspace{1mm}

 Such a (partial) gluing of the polytopes $\, M^{N}_{i} \, $ 
defines a mapping
$$\varphi :  \quad  \bigsqcup_{i} \, M^{N}_{i} 
\,\,\, \rightarrow \,\,\, \cal{T} \,\,\, , $$
which we shall call the partial gluing mapping.

\vspace{1mm}

 The topology of the space $\, \cal{T} \, $ is naturally defined by 
the continuity of the mapping $\, \varphi \, $ (a set in $\, \cal{T} \, $ is 
considered open if its full preimage in the 
space $\, \bigsqcup_{i} \, M^{N}_{i}$ is open).

\vspace{1mm}

 We also have a natural (continuous) mapping
$$g : \quad {\cal{T}} \,\,\, \rightarrow \,\,\, 
{\mathbb{T}}^{N} \,\,\, , $$
defined by the partition of the torus $\, \mathbb{T}^{N} \, $ 
(\ref{RazbienieTora}).

\vspace{1mm}

 Note that the points on the boundaries of $\, M^{N}_{i} \, $ that 
do not participate in the (partial) gluing are precisely the union of 
all sets of the form
$${\cal{P}}^{n}_{i} \, \times \, q^{N-n-1}_{i,p_{i}} \,\,\, , $$
where $\, {\cal{P}}^{n}_{i} \, $ is the interior of the polytope 
$\, P^{n}_{i} \, $, and $\, q^{N-n-1}_{i,p_{i}} \, $ is an arbitrary face of 
the polytope $\, Q^{N-n}_{i} \, $ of dimension $\, N - n - 1 \, $ 
(the remaining points on the boundaries of $\, M^{N}_{i} \, $ can be 
included in $(N-1)$ - dimensional faces that are orthogonal to $\, \xi$). 
Upon gluing these points to one another, the complex $\, \cal{T} \, $  
transforms into a manifold without boundary 
(namely, the torus $\, \mathbb{T}^{N}$). The complete boundary of the 
complex $\, \cal{T} \, $ is given by the union of the faces of
$\, M^{N}_{i} \, $, which take the form
\begin{equation}
\label{Pniqsi}
P^{n}_{i} \, \times \, q^{N-n-1}_{i,p_{i}} \,\,\, , 
\end{equation}
where $\, q^{N-n-1}_{i,p_{i}} \, $ is, in turn, an arbitrary face of the polytope 
$\, Q^{N-n}_{i} \, $ of dimension $\, N - n - 1 \, $ (we call a point 
$\, u \in \cal{T} \, $ an interior point of $\, \cal{T} \, $ if it has 
a spherical neighborhood $\, O_{\epsilon} (u) \subset \cal{T} \, $; 
otherwise, we call $\, u \, $ a boundary point of the complex $\, \cal{T} $).

\vspace{1mm}

 The union of the faces (\ref{Pniqsi}) (parallel to $\, \xi $) cannot 
divide the torus $\, \mathbb{T}^{N} \, $ into disconnected components. 
Indeed, a subspace $\, \mathbb{R}^{n} \subset \mathbb{T}^{N} \, $ of 
a generic direction $\, \xi \, $ does not intersect any of the sets 
(\ref{Pniqsi}) and is, at the same time, dense in $\, \mathbb{T}^{N} \, $. 
As can be seen from the structure of the complex $\, \cal{T} \, $, this 
implies its (linear and topological) connectedness. By the same reasoning, 
it can also be concluded that the set of interior points of $\, \cal{T} \, $ 
is connected.

\vspace{1mm}

 The partition of the torus $\, \mathbb{T}^{N} \, $ also defines a periodic 
partition of the space $\, \mathbb{R}^{N}$:
$$\mathbb{R}^{N} \,\,\, = \,\,\, \bigcup_{i,j} \, 
{\widehat{M}}^{N}_{i,j} \,\,\, , \quad \quad
(j = 1, \dots , \infty) \,\,\, , $$ 
such that each polytope $\, {\widehat{M}}^{N}_{i,j} \, $ maps to its 
corresponding polytope $\, M^{N}_{i} \, $ under the covering map 
$\, \mathbb{R}^{N} \rightarrow \mathbb{T}^{N} \, $. 

\vspace{1mm}

 By treating the collection of polytopes $\, \widehat{M}^{N}_{i,j} \, $ 
as abstract objects and performing a partial identification of their faces 
that are in contact within $\, \mathbb{R}^{N} \, $ (following the same scheme 
as that used for the set $\, \left\{ M^{N}_{i} \right\}$), we obtain a complex 
$\, \widehat{\cal{T}} \, $ that acts as a covering of $\, \cal{T} \, $:
$$\widehat{\cal{T}} \,\,\, \rightarrow \,\,\, \cal{T} \,\,\, , $$
analogous to the covering
$\, \mathbb{R}^{N} \rightarrow \mathbb{T}^{N} \, $. 

\vspace{1mm}

 As in the space $\, \mathbb{R}^{N} \, $, integer shifts are defined on the 
complex $\, \widehat{\cal{T}} \, $:
$$u \,\,\, \rightarrow \,\,\, u \, + \, {\bf N} \,\,\, , \quad \quad
{\bf N} \in \mathbb{Z}^{N} $$

 We also have a natural mapping (the mapping of the complete gluing)
$$\hat{g} : \quad \widehat{\cal{T}} \,\,\, \rightarrow \,\,\, 
\mathbb{R}^{N}  $$

 We shall say that points $\, u, \, u^{\prime} \in \widehat{\cal{T}} \, $ 
are at a distance $\, \epsilon \, $ from each other if
$$\left| \hat{g} (u) - \hat{g} (u^{\prime}) \right| 
\,\,\, = \,\,\, \epsilon $$

 Here, we define the full $\epsilon$ - neighborhood 
$\, O^{*}_{\epsilon} (u) \, $ of a point $\, u \in \widehat{\cal{T}} \, $
as the set of points $\, u^{\prime} \in \widehat{\cal{T}} \, $ whose distance 
from $\, u \, $ is less than $\, \epsilon \, $. Simultaneously, we define the 
$\epsilon$ - neighborhood $\, O_{\epsilon} (u) \, $ of a point 
$\, u \in \widehat{\cal{T}} \, $ as the (maximal) connected component of 
$\, O^{*}_{\epsilon} (u) \, $ that contains $\, u \, $ (the same applies 
to the complex $\, \cal{T} \, $ for sufficiently small $\, \epsilon$).

\vspace{1mm}

 As in the case of $\, \cal{T} \, $, we call a point 
$\, u \in \widehat{\cal{T}} \, $ an interior point of the complex 
$\, \widehat{\cal{T}} \, $ if, for sufficiently small $\, \epsilon \, $, 
its full neighborhood $\, O^{*}_{\epsilon} (u) \, $ is an open ($N$-dimensional) 
ball of radius $\, \epsilon \, $. The remaining points 
$\, u \in \widehat{\cal{T}} \, $ are called boundary points of 
$\, \widehat{\cal{T}} \, $ (the same applies to the complex $\, \cal{T} $). 
Clearly, for interior points $\, u \in \widehat{\cal{T}} \, $ 
(or $\, u \in \cal{T} \, $) and sufficiently small $\, \epsilon \, $, 
the sets $\, O^{*}_{\epsilon} (u) \, $ and $\, O_{\epsilon} (u) \, $ coincide.

\vspace{1mm}

 The boundary of the complex $\, \widehat{\cal{T}} \, $ is the union of an 
infinite set of polytopes
\begin{equation}
\label{GranitsavTAU}
\partial \, \widehat{\cal{T}} \,\,\, = \,\,\, \bigcup_{i,p_{i},j} \,
{\widehat{G}}^{N-1}_{i,p_{i},j} 
\end{equation}
$(j = 1, \dots , \infty)$, where each $\, {\widehat{G}}^{N-1}_{i,p_{i},j} \, $
is identical to one of the polytops (\ref{Pniqsi}).

 The mapping $\, \hat{g} \, $ maps the set 
$\, \partial \, \widehat{\cal{T}} \, $ into a periodic complex
\begin{equation}
\label{GranitsavRN}
\bigcup_{k} \, G^{N-1}_{k} \,\,\, \subset \,\,\, \mathbb{R}^{N} 
\end{equation}
$(k = 1, \dots , \infty)$, where each of the polytops $\, G^{N-1}_{k} \, $ 
also possesses the structure (\ref{Pniqsi}).

\vspace{1mm}

 It is easy to see that each point $\, {\bf X} \in \mathbb{R}^{N} \, $ 
has at most a finite number of preimages under the mapping $\, \hat{g} \, $.
It is also evident that the mapping $\, \hat{g} \, $ is one-to-one on the 
interior points of $\, \widehat{\cal{T}} \, $ (the same applies to the 
mapping $\, g: \, \cal{T} \rightarrow \mathbb{T}^{N}$).

\vspace{2mm}

 We shall call a function $\, F (u) \, $ smooth on $\, \cal{T} \, $ if, 
for any point $\, u \in \cal{T} \, $, there exists a neighborhood 
$\, O_{\epsilon} (u) \, $ such that the restriction
$$F \, \big|_{O_{\epsilon} (u)} $$
coincides with
$$g^{-1} \big( \widetilde{F} \big) \, \big|_{O_{\epsilon} (u)} $$
for some smooth function
$\, \widetilde{F} \left( {\bf X} \right) \, $ defined (everywhere) 
within the corresponding ball-shaped neighborhood
$$O_{\epsilon} \big( g (u) \big) \,\,\, \subset \,\,\, 
\mathbb{T}^{N} $$
(note here that the functions 
$\, \widetilde{F} \left( {\bf X} \right) \, $ 
and $\, \widetilde{F}^{\prime} \left( {\bf X} \right) \, $, defined 
for different points $\, u, u^{\prime} \in \cal{T} \, $ whose images 
$\, g (u) \, $ and $\, g (u^{\prime}) \, $ coincide, 
are not required to be identical).

\vspace{1mm}

 We will also use an analogous definition for smooth functions 
$\, \widehat{F} (u) \, $ on the complex $\, \widehat{\cal{T}} \, $.

\vspace{3mm}

\noindent
{\bf Definition.}

\vspace{1mm}

 We shall call a continuous embedding
$$\psi : \quad  \mathbb{R}^{n} \,\,\, \rightarrow \,\,\, 
\widehat{\cal{T}} $$
an affine embedding of $\, \mathbb{R}^{n} \, $ into 
$\, \widehat{\cal{T}} \, $ if the composition
$$\chi : \quad  \mathbb{R}^{n} \,\,\, \xrightarrow{\psi} \,\,\,
\widehat{\cal{T}} \,\,\, \xrightarrow{\hat g} \,\,\, \mathbb{R}^{N} $$
defines a standard affine embedding 
$\, \mathbb{R}^{n} \subset \mathbb{R}^{N} \, $ with direction $\, \xi \, $.

\vspace{3mm}

\noindent
{\bf Lemma 2.1}

\vspace{1mm}

 Let
$$\psi : \quad  \mathbb{R}^{n} \,\,\, \rightarrow \,\,\, 
\widehat{\cal{T}} $$
be an affine embedding, and let $\, \widehat{F} (u) \, $ be a smooth 
function on $\, \widehat{\cal{T}} \, $.

\vspace{1mm}

 Then
$$f ({\bf x}) \,\,\, = \,\,\, \widehat{F} \, 
\big|_{\mathbb{R}^{n} \subset \widehat{\cal{T}}} $$
is a smooth function on $\, \mathbb{R}^{n} \, $.

\vspace{2mm}

\noindent
Proof.

\vspace{1mm}

 Consider an arbitrary point $\, {\bf x} \in \mathbb{R}^{n} \, $
and its $\, \epsilon$ - neighborhood 
$\, O_{\epsilon} ({\bf x}) \subset \mathbb{R}^{n} \, $. 
Let $\, \epsilon \, $ be sufficiently small.

 By definition
$${\hat g} \Big( \psi \big( O_{\epsilon} ({\bf x}) \big) \Big) 
\,\,\, = \,\,\, \chi \big( O_{\epsilon} ({\bf x}) \big) 
\,\,\, = \,\,\, \chi \big( \mathbb{R}^{n} \big) \, \bigcap \,
O_{\epsilon} \big( \chi ({\bf x}) \big) \,\,\,\,\, , $$
where $\, O_{\epsilon} \big( \chi ({\bf x}) \big) \, $ is an 
$\, N$-dimensional $\, \epsilon$ - neighborhood of the point 
$\, \chi ({\bf x}) \, $ in $\, \mathbb{R}^{N} \, $.

\vspace{1mm}

 The set $\, \psi \big( O_{\epsilon} ({\bf x}) \big) \, $ is connected, 
belongs to 
$\, {\hat g}^{-1} \Big( O_{\epsilon} \big( \chi ({\bf x}) \big) \Big) \, $
and, thus, is contained in
$$O_{\epsilon} \big( \psi ({\bf x}) \big) \,\,\, \subset \,\,\, 
\widehat{\cal{T}} $$

\vspace{1mm}

 Then, by definition, we have (for sufficiently small $\, \epsilon$):
$$ \widehat{F} \, \big|_{O_{\epsilon} ({\bf x})} \,\,\, \equiv \,\,\,
\widehat{F} \, \big|_{\psi \left( O_{\epsilon} ({\bf x}) \right)} 
\,\,\, = \,\,\, \widetilde{F} \, 
\big|_{\chi \left( \mathbb{R}^{n} \right) \, \bigcap \,
O_{\epsilon} \left( \chi ({\bf x}) \right)} \,\,\, , $$
where $\, \widetilde{F} \left( {\bf X} \right) \, $ is a smooth 
function in the domain
$$O_{\epsilon} \big( \chi ({\bf x}) \big) \,\,\, \subset \,\,\, 
\mathbb{R}^{N} $$

 Thus, $\, f ({\bf x}) \, $ is smooth in the region 
$\, O_{\epsilon} ({\bf x}) \subset \mathbb{R}^{n} \, $.

\vspace{1mm}

\hfill{Lemma 2.1 is proved.}

\vspace{3mm}

 For simplicity, we will call the image of an affine embedding 
$\, \mathbb{R}^{n} \rightarrow \widehat{\cal{T}} \, $ 
an affine subspace of $\, \widehat{\cal{T}} \, $ (of dimension $n$). 
For a generic point $\, u \in \widehat{\cal{T}} \, $,
there is a unique affine subspace of dimension $\, n \, $ passing 
through $\, u \, $. For special points $\, u \in \widehat{\cal{T}} \, $ 
(those for which an affine subspace passing through them contains 
boundary points of $\, \widehat{\cal{T}} $), the number of such affine 
subspaces may be greater than 1. All affine subspaces 
$\, \mathbb{R}^{n} \subset \widehat{\cal{T}} \, $ that contain a given 
point $\, u \in \widehat{\cal{T}} \, $ are mapped to the same (affine) 
subspace $\, \mathbb{R}^{n} \subset \mathbb{R}^{N} \, $ under the 
mapping $\, g : \, \widehat{\cal{T}} \rightarrow \mathbb{R}^{N} \, $. 

\vspace{1mm}

 It should be noted that our definition of an affine subspace 
in $\, \widehat{\cal{T}} \, $ appears somewhat artificial (for instance, 
all such affine subspaces have the same direction $\, \xi$). 
Nevertheless, it will be convenient for us to use this term here.

\vspace{2mm}

 Through every point $\, u \in \widehat{\cal{T}} \, $, there passes 
at least one affine subspace $\, {\cal A} \subset \widehat{\cal{T}} \, $.
To show this, consider the image $\, \hat{g} (u) \in \mathbb{R}^{N} \, $
and the subspace $\, \mathbb{R}^{n} \subset \mathbb{R}^{N} \, $
of the direction $\, \xi \, $ that passes through $\, \hat{g} (u) \, $.

\vspace{1mm}

 If $\, \mathbb{R}^{n} \, $ does not intersect the set (\ref{GranitsavRN}), 
then the mapping $\, \hat{g} \, $ is one-to-one on it, and there exists 
a unique preimage $\, {\cal A} \subset \widehat{\cal{T}} \, $,
that contains $\, u \, $.

\vspace{1mm}

 Now suppose $\, u \notin \, \partial \widehat{\cal{T}} \, $,  yet the 
subspace $\, \mathbb{R}^{n} \, $ passing through $\, \hat{g} (u) \, $ 
intersects the set (\ref{GranitsavRN}). In this case, the intersection of 
$\, \mathbb{R}^{n} \, $ with the set (\ref{GranitsavRN}) presents an ambiguity 
regarding the choice of a preimage 
$\, {\hat g}^{-1} \left( {\bf X} \right) \, $. To resolve this, we choose 
a small vector $\, {\bf a} \in \mathbb{R}^{N} \, $ that is orthogonal to 
$\, \xi \, $ and such that all embeddings
\begin{equation}
\label{Rnlambda}
\mathbb{R}^{n} \, + \, \lambda \, {\bf a} \,\,\, , \quad \quad
\lambda \in (0, 1) 
\end{equation}
do not intersect with (\ref{GranitsavRN}). All embeddings (\ref{Rnlambda}) 
have unique preimages $\, {\cal A} (\lambda) \, $ in $\, \widehat{\cal{T}} \, $,
which define a continuous embedding 
$\, \mathbb{R}^{n} \rightarrow \widehat{\cal{T}} \, $ in the limit 
$\, \lambda \rightarrow 0 \, $. (This embedding is not unique and 
depends on the choice of the vector $\, {\bf a}$).

\vspace{1mm}

 Now let $\, u \in \, \partial \widehat{\cal{T}}$. If $\, u \, $ belongs 
to one of the faces $\, {\widehat{G}}^{N-1}_{i,p_{i},j} \, $ defined in 
(\ref{GranitsavTAU}) and does not belong to any face orthogonal to $\, \xi \, $, 
then there exists a unique polytope $\, {\widehat{M}}^{N}_{i,j}$ whose boundary 
contains $\, u \, $. In this case, the procedure described above can be used 
to construct $\, {\cal A} \ni u \, $, with the additional condition that the 
vector $\, {\bf a} \, $ is directed into the interior of the polytope 
$\, {\widehat{M}}^{N}_{i,j} \, $.

\vspace{1mm}

 Finally, if $\, u \, $ belongs to one of the faces in (\ref{GranitsavTAU}) and, 
at the same time, belongs to a face orthogonal to $\, \xi \, $, then there exist 
several polytopes $\, {\widehat{M}}^{N}_{i^{\prime},j^{\prime}}$ whose boundaries 
contain $\, u \, $. To construct $\, {\cal A} \ni u \, $ in this case, one can use 
the procedure described above, with the vector $\, {\bf a} \, $ directed into the 
interior of any one of these polytopes $\, {\widehat{M}}^{N}_{i^{\prime},j^{\prime}}$.

\vspace{3mm}

\noindent
{\bf Comment.}

 In the theory of quasicrystals, the passing of the subspace 
$\, \mathbb{R}^{n} (\lambda) \subset \mathbb{R}^{N} \, $ 
through the set (\ref{GranitsavRN}) corresponds to an instantaneous 
(local) change in the quasicrystalline structure within $\, \mathbb{R}^{n} \, $.
At the precise moment of intersection, the quasicrystalline structure in 
$\, \mathbb{R}^{n} (\lambda) \, $ is determined by an additional rule specific 
to the given model. The introduction of the complex $\, \widehat{\cal{T}} \, $
provides a convenient way to consider all possible variants simultaneously by 
examining various embeddings of $\, \mathbb{R}^{n} \, $ into 
$\, \widehat{\cal{T}} \, $. 

\vspace{3mm}

 We would now like to introduce additional general-position conditions 
on the direction $\, \xi \, $ and the corresponding partition of the torus 
$\, \mathbb{T}^{N} \, $ (\ref{RazbienieTora}).

\vspace{2mm}

 Namely, let us consider the set of all $\, (N-n-1)$ - dimensional faces of 
the polytopes $\, Q^{N-n}_{i}$. Let the (finite) set 
$\, \left\{ \eta_{j} \in G_{N, N-n-1} \right\} \, $ 
specify all possible directions of such faces in $\, \mathbb{R}^{N} \, $.

\vspace{2mm}

\noindent
(g2) We shall now require that none of the (hyper)planes
\begin{equation}
\label{xietaplane}
\xi \, \times \, \eta_{j} \,\,\, \in \,\,\, G_{N, N-1} 
\end{equation}
contain any integer vectors in $\, \mathbb{R}^{N} \, $.

\vspace{3mm}

\noindent
{\bf Proposition 2.1}

\vspace{1mm}

 Under the conditions stated above ((g1)–(g2)), any affine subspace 
$\, \mathbb{R}^{n} \subset \mathbb{R}^{N} \, $ with direction $\, \xi \, $ 
can intersect only a finite number of the polytopes $\, G^{N-1}_{k} \, $ 
that make up the complex (\ref{GranitsavRN}).

\vspace{2mm}

\noindent
Proof.

\vspace{1mm}

  Indeed, the intersection of $\, \mathbb{R}^{n} \, $ with an infinite 
number of $\, G^{N-1}_{k} \, $ would imply the existence of two identical 
polytopes $\, G^{N-1}_{k} \, $ and $\, G^{N-1}_{k^{\prime}} \, $ (differing 
by a shift by an integer vector 
$\, {\bf a} \in \mathbb{Z}^{N} \subset \mathbb{R}^{N} $)
that intersect $\, \mathbb{R}^{n} \, $. The latter, in turn, would imply 
the existence of an integer vector within the corresponding hyperplane 
(\ref{xietaplane}).

\vspace{1mm}

\hfill{Proposition 2.1 is proved.}

\vspace{3mm}

 The differences between quasicrystalline structures in two affine 
subspaces in $\, \widehat{\cal T} \, $
 \begin{equation}
\label{A1A2}
{\cal A}_{1} \,\,\, = \,\,\, \mathbb{R}^{n} \,\,\, \subset \,\,\, 
\widehat{\cal T} \,\,\, , \quad \quad
{\cal A}_{2} \,\,\, = \,\,\, \mathbb{R}^{n} \,\,\, \subset \,\,\, 
\widehat{\cal T} \,\,\, ,
\end{equation}
which pass through the same point $\, u \in \widehat{\cal T} \, $, 
arise because they pass through different polytopes 
$\, {\widehat{G}}^{N-1}_{i,p_{i},j} \, $ of the complex (\ref{GranitsavTAU}).
At the same time, they are mapped into the same subspace 
$\, \mathbb{R}^{n} \subset \mathbb{R}^{N} \, $ under the mapping 
$\, {\hat g} \, $, such that $\, {\hat g} \left( {\cal A}_{1} \right) \, $ 
and $\, {\hat g} \left( {\cal A}_{2} \right) \, $ intersect the same subset 
(${\cal G}$) of the faces $\, G^{N-1}_{k} \, $ from the set (\ref{GranitsavRN}). 
Differences between the quasicrystalline structures in 
$\, {\cal A}_{1} \, $ and $\, {\cal A}_{2} \, $ thus arise at their intersection 
with the set $\, {\hat g}^{-1} \left( {\cal G} \right) \, $. Proposition 2.1 
implies then the following proposition:

\vspace{3mm}

\noindent
{\bf Proposition 2.2}

\vspace{1mm}

 Under the stated general-position conditions:

\vspace{1mm}

\noindent
1) Any two affine subspaces (\ref{A1A2}) passing through the same point 
$\, u \in \widehat{\cal T} \, $ possess quasicrystalline structures that 
coincide everywhere outside a certain ball of finite radius;

\vspace{1mm}

\noindent
2) The restrictions of any smooth function $\, \widehat{F} (u) \, $ 
to any two affine subspaces (\ref{A1A2}) passing through the same point 
$\, u \in \widehat{\cal T} \, $ coincide everywhere outside a certain ball 
of finite radius.
 
\vspace{3mm}

\noindent
{\bf Definition.}

\vspace{1mm}

 We shall call an affine embedding 
$$\psi^{\prime} : \quad  \mathbb{R}^{n} \,\,\, \rightarrow \,\,\, 
\widehat{\cal{T}} $$
a shift of an affine embedding 
$$\psi : \quad  \mathbb{R}^{n} \,\,\, \rightarrow \,\,\, 
\widehat{\cal{T}} $$
(by a vector $\, {\bf a}$) if the composition
$$\chi^{\prime} : \quad  \mathbb{R}^{n} \,\,\, \xrightarrow{{\psi}^{\prime}} 
\,\,\, \widehat{\cal{T}} \,\,\, \xrightarrow{\hat g} \,\,\, \mathbb{R}^{N} $$
defines a shift of the (standard) affine embedding
$$\chi : \quad  \mathbb{R}^{n} \,\,\, \xrightarrow{\psi} \,\,\,
\widehat{\cal{T}} \,\,\, \xrightarrow{\hat g} \,\,\, \mathbb{R}^{N} $$
by the vector $\, {\bf a} \in \mathbb{R}^{N} \, $ 
(the same applies to affine subspaces).

\vspace{1mm}

 We shall call a family of affine embeddings
$$\psi (t) : \quad  \mathbb{R}^{n} \,\,\, \rightarrow \,\,\, 
\widehat{\cal{T}} $$
a continuous family of shifts of an affine embedding
$$\psi : \quad  \mathbb{R}^{n} \,\,\, \rightarrow \,\,\, 
\widehat{\cal{T}} \,\,\, , $$
if the composition
$$\chi (t) : \quad  \mathbb{R}^{n} \,\,\, \xrightarrow{\psi (t)} 
\,\,\, \widehat{\cal{T}} \,\,\, \xrightarrow{\hat g} \,\,\, \mathbb{R}^{N} $$
defines a continuous family of shifts of an affine embedding
$$\chi : \quad  \mathbb{R}^{n} \,\,\, \xrightarrow{\psi} \,\,\,
\widehat{\cal{T}} \,\,\, \xrightarrow{\hat g} \,\,\, \mathbb{R}^{N} $$
by vectors $\, {\bf a} (t) \in \mathbb{R}^{N} \, $
(the same applies to affine subspaces).

\vspace{3mm}

 When considering shifts of affine subspaces, we assume that the vector 
$\, {\bf a} \, $ is orthogonal to the direction $\, \xi \, $, and that 
points in different subspaces $\, {\cal A} (t) \subset \widehat{\cal{T}} \, $
are identified with one another via a shift in $\, \mathbb{R}^{N} \, $ 
following the mapping $\, \hat{g} \, $.
 
\vspace{3mm}

\noindent
{\bf Definition.}

\vspace{1mm}

 We shall call a smooth function $\, F (u) \, $ a smooth function of 
quasicrystalline type on $\, \cal{T} \, $ (denoted $\, F_{\rm QT}(u)$) 
if the function
$$\varphi^{-1} \big( F (u) \big) $$
is constant along the sections $\, Q^{N-n}_{i} \, $
on the set $\, \bigsqcup_{i} \, M^{N}_{i} \, $. 

\vspace{1mm}

 We shall call a function $\, \widehat{F} (u) \, $ a smooth function of 
quasicrystalline type on  $\, \widehat{\cal{T}} \, $ 
(denoted as $\, {\widehat{F}}_{\rm QT}(u)$) if it has the form
$$\pi^{-1} \big( F_{\rm QT} (u) \big) \,\,\, , $$
where $\, F_{\rm QT} (u) \, $ is a smooth function of quasicrystalline 
type on $\, \cal{T} \, $, and
$$\pi : \quad  \widehat{\cal{T}} \,\,\, \rightarrow \,\,\, \cal{T} $$
is the natural covering of $\, \cal{T} \, $ by $\, \widehat{\cal{T}} \, $.

\vspace{3mm}

 In this work, we are primarily interested in smooth functions 
$\, {\widehat{F}}_{\rm QT} (u) \, $ of the quasicrystalline type and 
their restrictions $\, f ({\bf x}) \, $ to affine subspaces 
$\, \mathbb{R}^{n} \subset \widehat{\cal{T}} \, $. A key feature of such 
functions $\, f ({\bf x}) \, $ is that, under parallel shifts of the affine 
subspace, their variations are local in nature and occur only when the 
subspace $\, \mathbb{R}^{n} (t) \subset \widehat{\cal{T}} \, $
intersects the set (\ref{GranitsavTAU}).

\vspace{1mm}

 As we have already mentioned, we shall also assume here that the 
functions $\, f ({\bf x}) \, $ can have only certain types of singularities 
(multiple saddle points or isolated minima and maxima). This, in particular, 
will allow us to consider their level surfaces 
$\, f ({\bf x}) = {\rm const} \, $ 
(whether singular or non-singular).

\vspace{3mm}

 Generic affine subspaces $\, {\cal A} \subset \widehat{\cal{T}} \, $
do not intersect the set (\ref{GranitsavTAU}). It is easy to see that for any 
bounded domain $\, \Omega \subset {\cal A} \, $, there exists in this case 
a sufficiently small $\, \epsilon > 0 \, $ such that all affine subspaces 
$\, {\cal A}^{\prime} \subset O^{*}_{\epsilon} \left( {\cal A} \right) 
\subset \widehat{\cal{T}} \, $
have no intersection with the set (\ref{GranitsavTAU}) within the domain 
$\, \Omega \, $. As a result, the restriction $\, f |_{\Omega} \, $
of a function $\, f ({\bf x}) \, $ - defined via a smooth function of 
quasicrystalline type - remains the same for all 
$\, {\cal A}^{\prime} \subset O^{*}_{\epsilon} \left( {\cal A} \right) \, $.

\vspace{1mm}

 Let us note here that, as before, 
$\, O^{*}_{\epsilon} \left( {\cal A} \right) \, $ 
represents the set of points $\, u \in \widehat{\cal{T}} \, $ whose images under 
the mapping $\, g \, $ are at a distance of less than $\, \epsilon \, $ from the 
set $\, g \left( {\cal A} \right) \, $. In the topology of the complex 
$\, \widehat{\cal{T}} \, $, the neighborhoods 
$\, O^{*}_{\epsilon} \left( {\cal A} \right) \, $ are open sets containing the 
subspace $\, {\cal A} \, $. 

\vspace{1mm}

 In the most general case (non-generic $\, {\cal A}$), for any bounded region 
$\, \Omega \subset {\cal A} \, $, there exists a sufficiently small neighborhood 
$\, O^{*}_{\epsilon} \left( {\cal A} \right) \subset \widehat{\cal{T}} \, $
such that any affine subspace 
$\, {\cal A}^{\prime} \subset O^{*}_{\epsilon} \left( {\cal A} \right) \, $
can intersect in $\, \Omega \, $ only those polytops 
$\, {\widehat{G}}^{N-1}_{i,p_{i},j} \, $ that lie in the 
full preimage $\, {\hat g}^{-1} \left( {\cal G} \right) \, $, where 
$${\cal G} \,\,\, = \,\,\, \bigcup \, {G^{N-1}_{k}}^{\prime} $$
is a collection of polytops (\ref{GranitsavRN}) that intersect
$\, {\hat g} \left( {\cal A} \right) \, $.

\vspace{1mm}

 Based on Propositions 2.1 and 2.2 (under the general-position conditions 
introduced above), we can formulate the following proposition.

\vspace{3mm}

\noindent
{\bf Proposition 2.3}

\vspace{1mm}

 For any affine subspace $\, {\cal A} \subset \widehat{\cal{T}} \, $, 
there exists a ball of finite radius $\, B^{n}_{r} \subset {\cal A} \, $ 
such that:

\vspace{1mm}

\noindent
1) For any ball $\, B^{n}_{R} \, $, there exists an $\, \epsilon$ - neighborhood 
$\, O^{*}_{\epsilon} \left( {\cal A} \right) \subset \widehat{\cal{T}} \, $
such that all affine subspaces 
$\, {\cal A}^{\prime} \subset O^{*}_{\epsilon} \left( {\cal A} \right) \, $
possess the same quasicrystalline structure within the region 
$\, B^{n}_{R} \big\backslash B^{n}_{r} \, $; 

\vspace{1mm}

\noindent
2) In particular, the values of a function $\, f ({\bf x}) \, $,
defined using any smooth function of quasicrystalline type on 
$\, \widehat{\cal{T}} \, $, remain the same within the region 
$\, B^{n}_{R} \big\backslash B^{n}_{r} \, $ for all affine subspaces 
$\, {\cal A}^{\prime} \subset O^{*}_{\epsilon} \left( {\cal A} \right) \, $.

\vspace{3mm}

 It is also easy to see that the presence of a ball $\, B^{n}_{r} \, $ 
(of radius $\, r \left( {\cal A} \right)$) pertains only to those subspaces 
$\, {\cal A} \subset \widehat{\cal{T}} \, $ that intersect the set 
(\ref{GranitsavTAU}). For generic subspaces $\, {\cal A} \, $, 
one may set $\, B^{n}_{r} = \emptyset \, $. 

\vspace{3mm}

 We now turn to the investigation of the Novikov problem, i.e. the problem 
of describing the global geometry of the level surfaces 
$\, f ({\bf x}) = {\rm const} \, $ for quasiperiodic functions 
$\, f ({\bf x}) \, $ of the B-type. We shall therefore assume 
that the function $\, f ({\bf x}) \, $ is the restriction of a certain smooth 
function of quasicrystalline type, $\, {\widehat{F}}_{\rm QT} (u) \, $,
to an affine subspace 
$\, {\cal A} = \mathbb{R}^{n} \subset \widehat{\cal{T}} \, $
within a complex $\, \widehat{\cal{T}} \, $ that corresponds to a certain
partition of the torus $\, \mathbb{T}^{N}$ (\ref{RazbienieTora}).
 
\vspace{1mm}

 In the Novikov problem for potentials of type A, it proves highly convenient 
to study the surfaces $\, f ({\bf x}) = {\rm const} \, $ simultaneously in all 
subspaces $\, \mathbb{R}^{n} \subset \mathbb{R}^{N} \, $ of a given direction 
$\, \xi \in G_{N,n} \, $. Here, we will also consider these surfaces across all 
affine subspaces $\, {\cal A} \subset \widehat{\cal{T}} \, $ for a fixed 
function $\, {\widehat{F}}_{\rm QT} (u) \, $. In this context, the level surfaces 
of the functions $\, f ({\bf x}) = c \, $ can be viewed as the intersections of 
the subspaces $\, {\cal A} \subset \widehat{\cal{T}} \, $ with the level sets 
$\, {\widehat{F}}_{\rm QT} (u) = c \, $.

\vspace{1mm}

 The sets $\, {\widehat{F}}_{\rm QT} (u) = c \, $ map to compact sets 
$\, {F}_{\rm QT} (u) = c \, $ in the complex $\, {\cal T} \, $
under the covering map $\, \pi : \, \widehat{\cal{T}} \rightarrow {\cal T} \, $.
In this process, the surfaces $\, f ({\bf x}) = c \, $ map to 
$\, (n-1)$ - dimensional layers, which are defined on the sets 
$\, {F}_{\rm QT} (u) = c \, $ by means of the direction $\, \xi \, $. 

\vspace{1mm}

 The primary objective of Novikov’s problem is to describe the global 
geometry of open (unbounded) level surfaces $\, f ({\bf x}) = {\rm const} \, $.
At the same time, it is also crucially important to describe the closed (bounded) 
level surfaces $\, f ({\bf x}) = {\rm const} \, $, which serve as the complement 
to the open level surfaces of $\, f ({\bf x}) \, $. It is easy to see that, 
for a generic direction $\, \xi \, $, the closed level surfaces 
$\, f ({\bf x}) = c \, $ map to compact leaves, while the open level surfaces 
map to non-compact (open) leaves on the set $\, {F}_{\rm QT} (u) = c \, $ 
under the mapping $\, \pi : \, \widehat{\cal{T}} \rightarrow {\cal T} \, $.

  The following lemmas play an important role in the study of the Novikov 
problem for potentials of type A (see \cite{DynMalNovUMN,BigQuas}). We also 
provide their proofs here for the case of potentials of type B (assuming all
the general-position conditions we adopted earlier are satisfied).

\vspace{3mm}

\noindent
{\bf Lemma 2.2}

\vspace{1mm}

 Let us consider a partition of the torus $\, \mathbb{T}^{N} \, $ 
(\ref{RazbienieTora}) and a smooth function $\, {F}_{\rm QT} (u) \, $
of quasicrystalline type defined on the associated complex $\, {\cal T} \, $.
Let $\, {\widehat{F}}_{\rm QT} (u) $ be the corresponding function of 
quasicrystalline type defined on $\, \widehat{\cal{T}} \, $, and let 
$\, f^{\cal A} ({\bf x}) \, $ represent the restrictions of 
$\, {\widehat{F}}_{\rm QT} (u) $ to all possible affine subspaces 
$\, {\cal A} \subset \widehat{\cal{T}} \, $.
 
\vspace{1mm}
 
 Let the value $\, c \, $ be such that all connected components of the 
level surfaces $\, f^{\cal A} ({\bf x}) = c \, $ (in all subspaces 
$\, {\cal A} \subset \widehat{\cal{T}} $) are closed.

\vspace{1mm}

 Then the diameters of all components of the level surfaces 
$\, f^{\cal A} ({\bf x}) = c \, $ (in all subspaces 
$\, {\cal A} \subset \widehat{\cal{T}} $) are bounded by a single 
constant $\, D > 0 \, $. 

\vspace{2mm}

\noindent
Proof.

\vspace{1mm}

 Let $\, u \in \widehat{\cal{T}} \, $ be an arbitrary point belonging to 
the set $\, {\widehat{F}}_{\rm QT} = c \, $, and let 
$\, {\cal A} \in \widehat{\cal{T}} \, $ be an affine subspace in 
$\, \widehat{\cal{T}} \, $ that contains $\, u \, $. Let 
$\, L_{u} \subset {\cal A} \, $ be a (singular or non-singular) level surface 
$\, f^{\cal A} ({\bf x}) = c \, $ that contains $\, u \, $. 

 Consider an (open) ball $\, B^{n}_{R} \subset {\cal A} \, $ from 
Proposition 2.3, which contains $\, L_{u} \, $. If $\, L_{u} \, $ intersects 
the ball $\, B^{n}_{r} \subset B^{n}_{R}  \, $, we increase the radius 
$\, R \, $, if necessary, so that the region $\, B^{n}_{R} \, $
also contains all connected components of the set 
$\, f^{\cal A} ({\bf x}) = c \, $ that intersect $\, B^{n}_{r} \, $.

 According to Proposition 2.3, there exists a neighborhood
$\, O^{*}_{\epsilon} \left( B^{n}_{R} \right) \subset \widehat{\cal{T}} \, $ 
containing all connected components
$\, f^{{\cal A}^{\prime}} ({\bf x}) = c \, $ (in all subspaces
$\, {\cal A}^{\prime} \subset \widehat{\cal{T}} \, $  intersecting
$\, O^{*}_{\epsilon} \left( B^{n}_{R} \right)$) intersecting the neighborhood
$\, O^{*}_{\epsilon} (u) \subset \widehat{\cal{T}} \, $. 

 For sufficiently small $\, \epsilon \, $, the neighborhood 
$\, O^{*}_{\epsilon} (u) \, $ is mapped one-to-one onto its image
$$O^{*}_{\epsilon} \left( \pi (u) \right) \,\,\, = \,\,\,  
\pi \left( O^{*}_{\epsilon} (u) \right) \,\,\, \subset \,\,\, {\cal T} $$
under the mapping $\, \pi : \, \widehat{\cal{T}} \rightarrow {\cal T} \, $. 

 The set $\, O^{*}_{\epsilon} \left( \pi (u) \right) \, $ is an open 
neighborhood of the point $\, \pi (u) \in {\cal T} \, $ with the property that 
all fibers $\, f^{{\cal A}^{\prime}} ({\bf x}) = c \, $ on the set 
$\, {F}_{\rm QT} = c \, $ that pass through 
$\, O^{*}_{\epsilon} \left( \pi (u) \right) \, $ are compact, and their preimages 
in $\, \widehat{\cal{T}} \, $ have diameters bounded by the constant $\, 2 R \, $. 

 The union of the sets $\, O^{*}_{\epsilon} \left( \pi (u) \right) \, $
for all $\, u \, $ belonging to the set $\, {\widehat{F}}_{\rm QT} (u) = c \, $
defines an open covering of a compact set
$$\left\{ u \in {\cal T} : \,\,\, {F}_{\rm QT} (u) = c \right\} $$

 By choosing a finite subcover of this cover, we obtain a proof of the lemma.

\hfill{Lemma 2.2 is proved.}

\vspace{3mm}

 Under the conditions of Lemma 2.2, it can be stated that in each of the 
affine subspaces $\, {\cal A} \subset \widehat{\cal T} \, $, one of the 
following two situations occurs.

\vspace{2mm}

\noindent
1) Situation $\, A_{-} \, $.

\begin{figure}[t]
\begin{center}
\includegraphics[width=\linewidth]{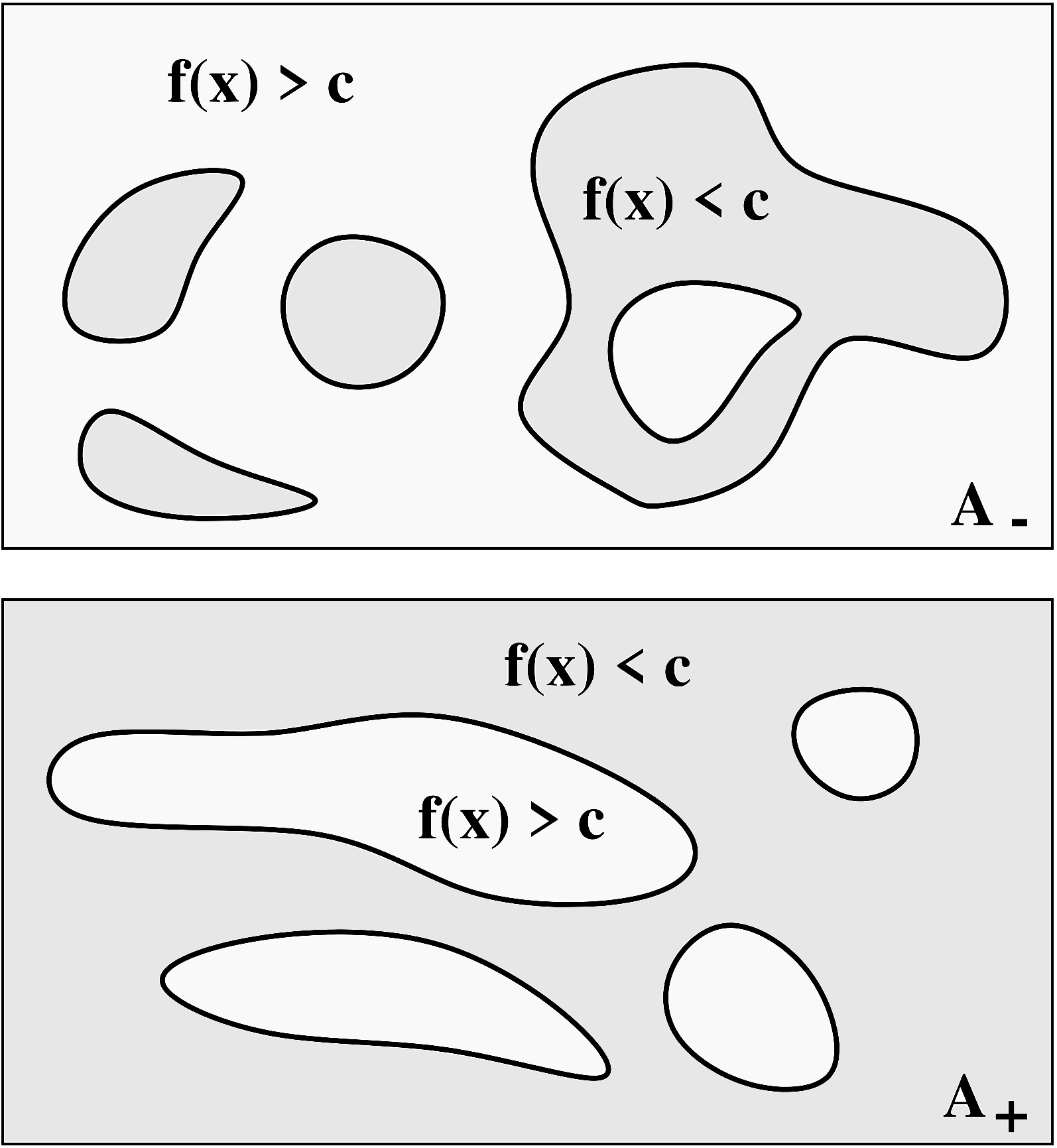}
\end{center}
\caption{Situations $\, A_{-} \, $ and $\, A_{+} \, $ in the space 
$\, \mathbb{R}^{n} \, $.}
\label{AMinusPlus}
\end{figure}

 The set $\, f^{\cal A} ({\bf x}) > c \, $ has a single unbounded connected 
component. The sizes of all other (connected) components of the sets 
$\, f^{\cal A} ({\bf x}) < c \, $ and $\, f^{\cal A} ({\bf x}) > c \, $
are bounded by the constant $\, D \, $ (Fig. \ref{AMinusPlus}).

\vspace{2mm}

\noindent
2) Situation $\, A_{+} \, $.

 The set $\, f^{\cal A} ({\bf x}) < c \, $ has a single unbounded connected 
component. The sizes of all other (connected) components of the sets 
$\, f^{\cal A} ({\bf x}) < c \, $ and $\, f^{\cal A} ({\bf x}) > c \, $
are bounded by the constant $\, D \, $ (Fig. \ref{AMinusPlus}).

\vspace{3mm}

 As follows from Proposition 2.3, under the conditions of Lemma 2.2, 
the situations ($A_{-}$ or $A_{+}$) that arise in all subspaces 
$\, {\cal A} \subset \widehat{\cal T} \, $ passing through any given point 
$\, u \in \widehat{\cal T} \, $ must coincide (the values of the functions
$\, f^{\cal A} ({\bf x}) \, $ in all such subspaces $\, {\cal A} \, $ are 
identical everywhere outside the ball $\, B^{n}_{r} (u)$). As can be seen 
also, any such subspace $\, {\cal A} \subset \widehat{\cal T} \, $
contains interior points $\, u \in \widehat{\cal T} \, $
(which do not belong to the set (\ref{GranitsavTAU})).

 It is also easy to see that the images of $\, {\cal A} \, $ and 
$\, {\cal A}^{\prime} \, $ under the mapping 
$\, {\hat g} : \,\, \widehat{\cal T} \rightarrow \mathbb{R}^{N} \, $
coincide if and only if $\, {\cal A} \, $ and $\, {\cal A}^{\prime} \, $ 
share common interior points $\, u \in \widehat{\cal T} \, $.

 Using the mapping $\, {\hat g} \, $, we can thus uniquely associate the 
situation $\, A_{-} \, $ or $\, A_{+} \, $ with each subspace 
$\, \mathbb{R}^{n} \subset \mathbb{R}^{N} \, $ of direction $\, \xi \, $ 
(provided the conditions of Lemma 2.2 are satisfied).

\vspace{3mm}

\noindent
{\bf Proposition 2.4}

\vspace{1mm}

 Under the conditions of Lemma 2.2, all affine subspaces 
$\, {\cal A} \subset \widehat{\cal T} \, $ correspond the same 
situation ($A_{-} \, $ or $\, A_{+}$).

\vspace{2mm}

\noindent
Proof.

\vspace{1mm}

 Let
$$W_{-} \subset \mathbb{R}^{N} \big/ \mathbb{R}^{n}
\, = \, \mathbb{R}^{N-n} \quad  \text{\rm and} \quad 
W_{+} \subset \mathbb{R}^{N} \big/ \mathbb{R}^{n}
\, = \, \mathbb{R}^{N-n} $$
be the sets of subspaces $\, \mathbb{R}^{n} \subset \mathbb{R}^{N} \, $ 
of directions $\, \xi \, $ corresponding to the situations 
$\, A_{-} \, $ and $\, A_{+} \, $, respectively. Let us show that both sets, 
$\, W_{-} \, $ and $\, W_{+} \, $, are open.

 Indeed, suppose, for example, that there exists a subspace 
$\, \mathbb{R}^{n} \subset \mathbb{R}^{N} \, $
and some preimage $\, {\cal A} \, $ of this subspace in 
$\, \widehat{\cal T} \, $ where we have the situation $\, A_{-} \, $.
This implies, in particular, that for a sufficiently large $\, R \, $, 
the set $\, B^{n}_{R} \big\backslash B^{n}_{r} \, $ contains a connected 
component of the set $\, f^{\cal A} ({\bf x}) > c \, $ with size greater 
than $\, D \, $. 

 According to Proposition 2.3, this same situation persists within a certain 
neighborhood 
$\, O^{*}_{\epsilon} \left( {\cal A} \right) \subset \widehat{\cal T} \, $,
which implies that the situation $\, A_{-} \, $ holds for all subspaces 
$\, {\cal A}^{\prime} \subset  O^{*}_{\epsilon} \left( {\cal A} \right) \, $.
The latter means that all subspaces 
$\, \mathbb{R}^{n} \subset \mathbb{R}^{N} \, $
lying within an $\, \epsilon$ - neighborhood of our chosen subspace 
$\, \mathbb{R}^{n} \, $ belong to $\, W_{-} \, $. 

 Thus, we can see that every point of $\, W_{-} \, $ is contained in 
$\, W_{-} \, $ along with some $\, \epsilon$ - neighborhood (the same applies 
to $\, W_{+}$). From the connectedness of the space
$$\mathbb{R}^{N-n} \,\,\, = \,\,\, \mathbb{R}^{N} \big/ \mathbb{R}^{n} 
\,\,\, , $$
we then obtain the proof of the proposition.

\vspace{1mm}

\hfill{Proposition 2.4 is proved.}

\vspace{3mm}

\noindent
{\bf Lemma 2.3}

\vspace{1mm}

 Let $\, \xi \in G_{N,n} \, $ satisfy the general position conditions (g1–g2), 
and let (\ref{RazbienieTora}) be the partition of the torus associated with 
the direction $\, \xi \, $.

\vspace{1mm}

 Let $\, F_{\rm QT} (u) \, $ be a smooth function of quasicrystalline type 
on $\, {\cal T} \, $, and let $\, {\widehat{F}}_{\rm QT} (u) \, $ be the 
corresponding function of quasicrystalline type on the complex 
$\, \widehat{\cal T} \, $.

\vspace{1mm}

 Assume that all the restrictions 
\begin{equation}
\label{fcalA}
f^{\cal A} ({\bf x}) \,\,\, = \,\,\, {\widehat{F}}_{\rm QT}
\big|_{{\cal A} \subset \widehat{\cal T}} 
\end{equation}
admit only certain types of singularities: 
multiple saddle points and isolated minima or maxima.

\vspace{1mm}

 Then the values $\, c \, $ for which open level surfaces
$$f^{\cal A} ({\bf x}) \,\,\, = c \,\,\, $$
arise in at least one of the subspaces 
$\, {\cal A} \subset \widehat{\cal T} \, $ form a closed interval 
$\, \left[ c_{1} , \, c_{2} \right] \, $.

\vspace{2mm}

\noindent
Proof.

\vspace{1mm}

 Indeed, if for a given $\, c_{0} \, $ all components of the level sets 
$\,  f^{\cal A} ({\bf x}) = c_{0} \, $ (within all subspaces 
$\, {\cal A} \subset \widehat{\cal T}$) are closed, then the same property 
holds for either all values $\, c < c_{0} \, $ or all values $\, c > c_{0} \, $
(depending on whether the subspaces $\, {\cal A} \subset \widehat{\cal T} \, $
correspond to the situation $\, A_{-} \, $ or $\, A_{+} $).

 Thus, the set of values $\, c \, $ that correspond to the emergence of 
unbounded components of the level set $\,  f^{\cal A} ({\bf x}) = c \, $
for at least one $\, {\cal A} \subset \widehat{\cal T} \, $ 
forms a connected interval.

\vspace{1mm}

 To prove that this interval is closed, one can show that its complement - 
that is, the set of values $\, c \, $ for which all level surfaces 
$\,  f^{\cal A} ({\bf x}) = c \, $ 
(for all $\, {\cal A} \subset \widehat{\cal T}$) are closed - is an open set.

\vspace{1mm}

 Let (without loss of generality) the level $\, c \, $ correspond to the 
situation $\, A_{-} \, $ in all subspaces 
$\, {\cal A} \subset \widehat{\cal T} \, $. 
This situation, evidently, also holds for all $\, c^{\prime} < c \, $. 
Now, let $\, c^{\prime} > c \, $. 

\vspace{1mm}

 As we have said, we assume that all restrictions (\ref{fcalA}) admit 
only certain types of singularities: multiple saddle points and isolated 
minima or maxima. Within any bounded domain $\, \Omega^{n} \subset {\cal A} \, $,
the value of $\, \big| \nabla f^{\cal A} ({\bf x}) \big| \, $ is bounded from 
below by a positive value, except within (arbitrarily small) neighborhoods of 
the singular points of $\, f^{\cal A} ({\bf x}) \, $. 

\vspace{1mm}

 It can then be seen that for any $\, {\cal A} \,$, any ball 
$\, B^{n}_{R} \subset {\cal A} \,$, and any $\, \epsilon > 0 \,$, 
there exists a $\, \delta > 0 \,$ such that any connected component 
of the set $\, f^{\cal A} ({\bf x}) < c^{\prime} \,$ that intersects 
$\, B^{n}_{R} \,$ is contained within the ball 
$\, B^{n}_{R + D(c) + \epsilon} \,$ (concentric with $\, B^{n}_{R}$) 
of radius $\, R + D(c) + \epsilon \,$, 
provided that $\, 0 < c^{\prime} - c < \delta \,$.

\vspace{1mm}

 A specific feature of functions of the quasicrystalline type, 
$\, {\widehat{F}}_{\rm QT} (u) \, $, is that any subspace 
$\, {\cal A} \subset \widehat{\cal T} \, $ can be covered by balls 
$\, B^{n}_{R} \, $ (of a sufficiently large radius $\, R$) such that, 
within each corresponding ball $\, B^{n}_{R + D(c) + \epsilon} \, $,
the function $\, f^{\cal A} ({\bf x}) \, $ (expressed in local coordinates) 
must belong to a fixed finite set of functions
$\, \left\{ \varphi_{i} ({\bf x}) \right\} \, $ - the same set for all 
$\, {\cal A} \subset \widehat{\cal T} \, $ - defined on the ball 
$\, B^{n}_{R + D(c) + \epsilon} \, $ (centered at the origin).

\vspace{1mm}

 From the above, we therefore conclude that the situation $\, A_{-} \, $
(for all $\, {\cal A} \subset \widehat{\cal T} \, $ simultaneously) is 
stable with respect to small variations in the level $\, c \, $ 
(the same applies to the situation $\, A_{+}$).

\vspace{1mm}

 It should also be noted that the emergence of situation $\, A_{-} \, $ near 
the value $\, F^{\rm min}_{\rm QT} \, $ and situation $\, A_{+} \, $ near 
$\, F^{\rm max}_{\rm QT} \, $ implies the existence of a non-empty interval 
$\, \left[ c_{1} , \, c_{2} \right] \, $ that separates these two situations.

\vspace{1mm}

\hfill{Lemma 2.3 is proved.}

\vspace{3mm}

 It should be noted here that Lemma 2.3 does not assert the existence of 
unbounded level components $\, f^{\cal A} ({\bf x}) = c \, $ for all 
$\, {\cal A} \subset \widehat{\cal T} \, $ when 
$\, c \in  \left[ c_{1} , \, c_{2} \right] \, $. 
However, the following lemma holds.

\vspace{3mm}

\noindent
{\bf Lemma 2.4}

\vspace{1mm}

 Under the conditions of the previous lemma, for values 
$\, c \in  \left[ c_{1} , \, c_{2} \right] \, $, each subspace 
$\, {\cal A} \subset \widehat{\cal T} \, $ contains either unbounded 
components of the surface $\, f^{\cal A} ({\bf x}) = c \, $
or its components of arbitrarily large (finite) size (or both).

\vspace{2mm}

\noindent
Proof.

\vspace{1mm}

 Let $\, c \in  \left[ c_{1} , \, c_{2} \right] \, $, 
and let $\, {\cal A}_{0} \subset \widehat{\cal T} \, $ be an affine 
subspace containing only those components of the surface 
$\, f^{{\cal A}_{0}} ({\bf x}) = c \, $ that have a diameter of at 
most $\, D \, $ (for some $\, D$).

 Let $\, {\cal A}^{\prime} \subset \widehat{\cal T} \, $ be a subspace 
containing an unbounded connected component of the surface 
$\, f^{\cal A^{\prime}} ({\bf x}) = c \, $. Consider a finite (connected) 
part $\, S \, $ of this component with diameter $\, > D \, $.  If the 
subspace $\, {\cal A}^{\prime} \, $  contains the ball $\, B^{n}_{r} \, $
from Proposition 2.3 (i.e., it is not a generic subspace), we may assume, 
without loss of generality, that $\, S \, $ does not intersect 
$\, B^{n}_{r} \, $. 

 Then there exists a neighborhood 
$\, O^{*}_{\epsilon} \left( {\cal A}^{\prime} \right) \, $ such that all 
subspaces $\, {\cal A} \, $ lying within 
$\, O^{*}_{\epsilon} \left( {\cal A}^{\prime} \right) \, $
contain a segment $\, S \, $ at the level $\, f^{\cal A} ({\bf x}) = c \, $.

 Under our general-position assumptions regarding the direction $\, \xi \, $, 
however, some integer shift of the subspace $\, {\cal A}_{0} \, $ must 
lie within $\, O^{*}_{\epsilon} \left( {\cal A}^{\prime} \right) \, $. 
This leads to a contradiction.

\vspace{1mm}

\hfill{Lemma 2.4 is proved.}

\vspace{3mm}

\noindent
{\bf Comment.}

 As shown in \cite{BigQuas}, for A-type potentials, the absence of unbounded 
components $\, f^{\cal A} ({\bf x}) = c \, $ within certain subspaces 
$\, {\cal A} \subset \widehat{\cal T} \, $ for 
$\, c \in  \left[ c_{1} , \, c_{2} \right] \, $ can occur only when 
$\, c = c_{1} \, $ or $\, c = c_{2} \, $. As we shall see below, 
for B-type potentials, this situation is more common.

\vspace{3mm}

 As we have already said, the statements given above also have analogues 
for potentials of type A. Nevertheless, potentials of types A and B also 
exhibit significant differences, which we present in the next chapter.

\section{The Novikov problem and the distinctions between potentials 
of types A and B}
\setcounter{equation}{0}

 Based on the representation (\ref{fFConnection}), it can be seen that, 
for both A-type and B-type potentials, the spectrum of the functions 
$\, f ({\bf x}) \, $ is determined by the spectrum of the function 
$\, F \left( {\bf X} \right) \, $. In both cases, the Fourier harmonics 
of the functions $\, f ({\bf x}) \, $ correspond to the wave vectors
$${\bf k}  \left( m^{1}, \dots , m^{N} \right) \,\,\, = \,\,\, 
m^{1} {\bf e}_{1} \, + \, \dots \, + \, m^{N} {\bf e}_{N} $$
($(m^{1}, \dots , m^{N}) \in \mathbb{Z}^{N}$), where the vectors
$\, \left\{ {\bf e}_{j} \in \mathbb{R}^{n} \right\} \, $ 
are linearly independent over the field of rational numbers.

\vspace{1mm}

 For functions $\, f ({\bf x}) \, $  of type A, one may assume 
(provided the function $\,  F \left( {\bf X} \right) \, $ is sufficiently 
smooth) that the decay of the Fourier coefficients of $\, f ({\bf x}) \, $
is determined by the magnitude of $\, | {\bf m} | \, $ and is quite rapid 
(for example, faster than any power of $\, | {\bf m} |$).

\vspace{1mm}

 In the case of B-type functions, however, the Fourier coefficients of the 
functions $\, F \left( {\bf X} \right) \, $ decay significantly more slowly 
(especially for wave vectors nearly orthogonal to the direction $\, \xi$). 
Consequently, the set of Fourier coefficients of $\, f ({\bf x}) \, $
includes, for instance, significant long-wavelength harmonics corresponding 
to relatively large values of $\, | {\bf m} | \, $
(see e.g. \cite{KaluginKitaevLevitov2}). Overall, the decay of
the Fourier coefficients in this case is associated more with an increase 
in the value of $\, | {\bf k} | \, $ than with an increase in 
$\, | {\bf m} | \, $.

\vspace{2mm}

 In addition to the differences in the behavior of Fourier harmonics, 
one can also point to distinct geometric differences between quasiperiodic 
functions of type A and type B. Indeed, the values (the graph) of a 
quasiperiodic function of type B in the space $\, \mathbb{R}^{n} \, $ 
consist of a repetition of its graphs over the elements of a tiling of 
$\, \mathbb{R}^{n} \, $, where only a finite number of distinct elements 
exist. Consequently, the singular points of such a function occur only on 
a finite number of its level sets, $\, f ({\bf x}) = c_{i} \, $.
This distribution of singular level sets differs sharply from their 
distribution for quasiperiodic functions of type A, which was 
studied in detail in the work \cite{GusseinZade1}.

\vspace{2mm}

 Significant differences between quasiperiodic functions of type B and 
those of type A also arise in the context of the Novikov problem. It 
should be noted that many results concerning the (both primary and generalized)
Novikov problem for potentials of type A are substantially based on the 
smoothness of the function $\, F \left( {\bf X} \right) \, $ that defines 
the functions $\, f ({\bf x}) \, $. Consequently, despite the results presented 
in the previous section, distinctions between functions of type B and type A 
manifest quite frequently here.

\vspace{1mm}

 One of the significant aspects of the Novikov problem is, in particular, 
finding the level sets of a function $\, f ({\bf x}) \, $ that contain open 
level surfaces of one type or another. For (smooth) quasiperiodic functions 
of type B, the following proposition can be easily formulated.

\vspace{3mm}

\noindent
{\bf Proposition 3.1}

\vspace{1mm}

 Any non-singular open level surface
\begin{equation}
\label{LevelSurf}
f ({\bf x}) \,\,\, = \,\,\, c 
\end{equation}
of a smooth quasiperiodic function $\, f ({\bf x}) \, $ of type B is stable
with respect to small variations in the level $\, c \, $. 

\vspace{2mm}

\noindent
Proof.

\vspace{1mm}

 Indeed, any non-singular level surface (\ref{LevelSurf}) for a B-type 
potential $\, f ({\bf x}) \, $ can be divided into parts belonging to 
different cells of the covering of $\mathbb{R}^{n} \, $. Among these parts, 
there are only a finite number of distinct types, and each one is non-singular. 
Thus, there exists a common interval 
$\, c \in \left( {\tilde c}_{1}, \, {\tilde c}_{2} \right) \, $
where each of these parts remains non-singular, and the 
scheme for connecting these parts to one another is preserved. 

\hfill{Proposition 3.1 is proved.}

\vspace{3mm}

 In fact, it follows from Proposition 3.1 that any open non-singular 
component of a level surface (\ref{LevelSurf}) of a quasiperiodic function 
of type B remains stable within a certain interval 
$\, c \in \left( {\tilde c}_{1}, \, {\tilde c}_{2} \right) \, $,
and becomes singular at its boundaries  (possibly by merging 
into a larger singular conglomerate).

\vspace{1mm}

 Proposition 3.1 demonstrates a clear distinction between quasiperiodic 
functions of type B and those of type A. Thus, for functions of type A, 
already in the case $\, n = 2 \, $, $ \, N = 3 \, $, the nonsingular open 
trajectories of the Tsarev or Dynnikov type (playing an important role in 
the theory of galvanomagnetic phenomena) exist at only one energy level.

\vspace{2mm}

 As another example, we consider a quasiperiodic function of type B that 
corresponds to the Penrose rhombic tiling (P3) and possesses non-singular open 
level lines within a finite interval of values 
$\, c \in \left( c_{1}, \, c_{2} \right) \, $. As was shown in \cite{Sirius}, 
quasiperiodic functions of type A in the plane (with an arbitrary number of 
quasiperiods) that exhibit dihedral symmetry can have open level lines only at 
a single specific level, $\, c = c_{0} \, $. The example presented below thus 
illustrates yet another distinction between functions of type A and type B.

\begin{figure}[t]
\begin{center}
\includegraphics[width=\linewidth]{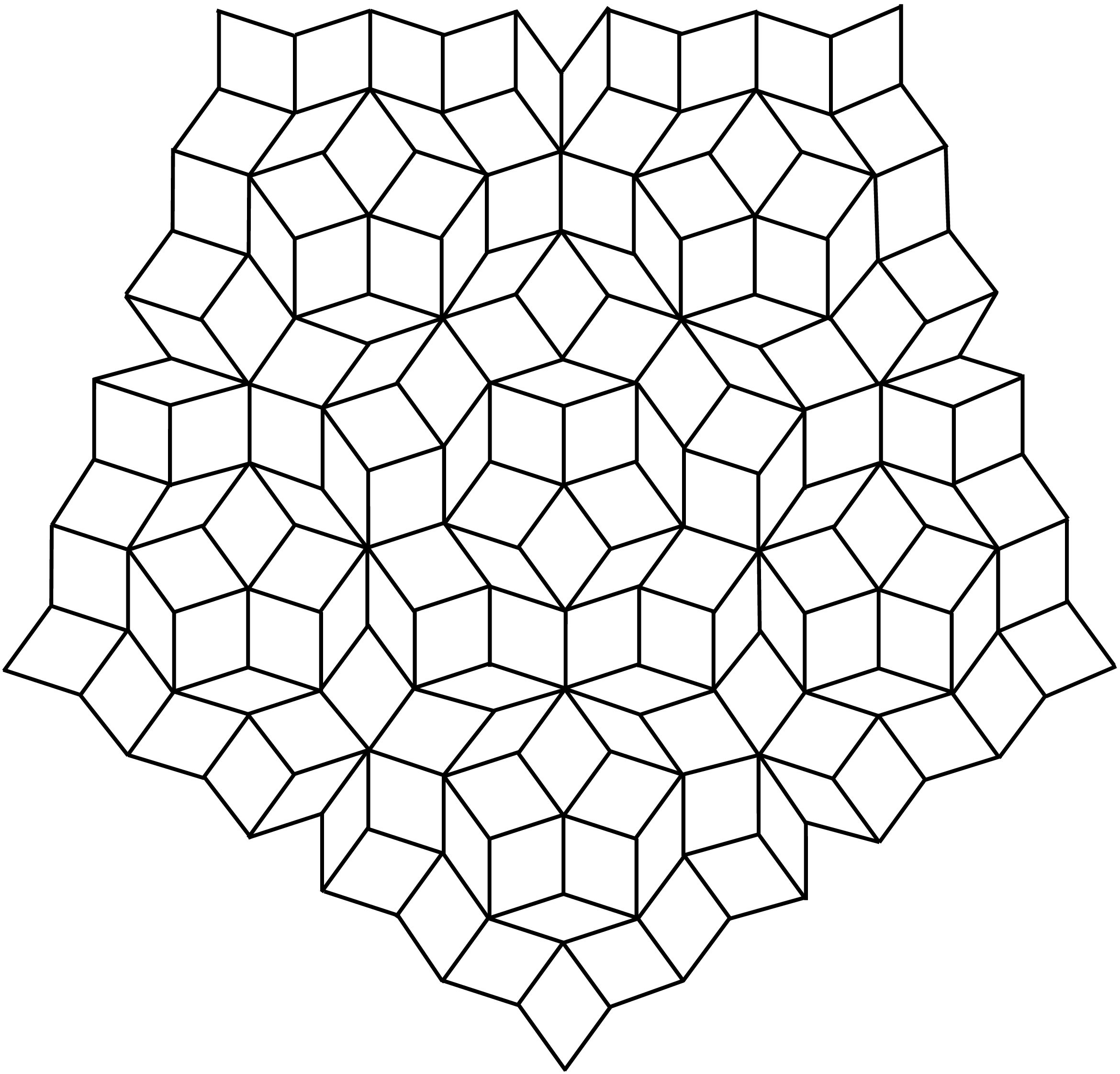}
\end{center}
\caption{A Penrose P3 tiling, which possesses exact 5-fold symmetry.}
\label{Penrose}
\end{figure}

 It is well known that Penrose rhombic tilings are formed by two types of 
rhombi: ``thick'' and ``thin'' rhombi (with acute angles of $72^{\circ}$ 
and $36^{\circ}$, respectively). Among all such tilings, one can specifically 
distinguish those that possess exact rotational symmetry in 
$\, \mathbb{R}^{2} \, $ (Fig. \ref{Penrose}).

 At the same time, the Penrose tilings can be defined via the section method for an 
embedding $\, \mathbb{R}^{2} \subset \mathbb{R}^{5} \, $ (see \cite{OgDunKatz}).

 The embedding $\, \mathbb{R}^{2} \subset \mathbb{R}^{5} \, $ is not completely 
irrational and is contained within the integral plane 
$\, \mathbb{R}^{4} \subset \mathbb{R}^{5} \, $ that is orthogonal to the vector
$${\bf e} \,\,\, = \,\,\, {\bf e}_{1} \, + \, {\bf e}_{2} \, + \,
{\bf e}_{3} \, + \, {\bf e}_{4} \, + \, {\bf e}_{5} $$

 Thus, functions $\, f (x, y) \, $ corresponding to the Penrose tiling are 
quasiperiodic functions with 4 quasiperiods.
 
\vspace{1mm}

\begin{figure}[t]
\begin{center}
\includegraphics[width=0.9\linewidth]{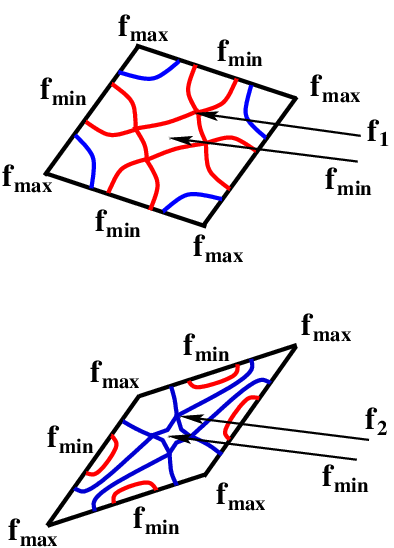}
\end{center}
\caption{Level lines and singular points of the function $\, f (x, y) \, $ 
on the ``thick'' and ``thin'' rhombi in the Penrose tiling.}
\label{PenroseFunctions}
\end{figure}

 To define the function $\, f (x, y) \, $ that we need, it suffices to specify 
its values on the ``thick'' and ``thin'' rhombi. Our function $\, f (x, y) \, $ 
will exhibit the symmetry of the corresponding rhombus (on each of the rhombi), 
and its level lines will display a similar geometry on both rhombi 
(Fig. \ref{PenroseFunctions}).

  The function $\, f (x, y) \, $ attains the same maximum value 
$\, f_{\rm max} \, $ at the vertices of both rhombi, and the same minimum 
value $\, f_{\rm min} \, $ at the midpoints of their sides. The function 
$\, f (x, y) \, $ has 4 saddle-type critical points at the level $\, f_{1} \, $
inside the ``thick'' rhombus, and 4 saddle-type critical points at the level 
$\, f_{2} > f_{1} \, $ inside the ``thin'' rhombus. Furthermore, 
$\, f (x, y) \, $ takes the value $\, f_{\rm min} \, $ at the centers of 
the ``thick'' and the ``thin'' rhombi (Fig. \ref{PenroseFunctions}).

 The level lines of the function $\, f (x, y) \, $ meet the boundaries of the 
rhombi at a $90^{\circ}$ angle, and the behavior of $\, f (x, y) \, $ near these 
boundaries ensures any desired degree of smoothness in the rhombic tiling of the 
space $\, \mathbb{R}^{2} \, $. It can be also verified that the function 
$\, f (x, y) \, $ is defined by a smooth, periodic function 
$\, \widehat{F} (u) \, $ of a quasicrystalline type on a complex 
$\, \widehat{\cal T} \, $ (which corresponds to a certain refinement 
of a partition of the torus presented in \cite{OgDunKatz}).

\vspace{2mm}

\begin{figure}[t]
\begin{center}
\includegraphics[width=\linewidth]{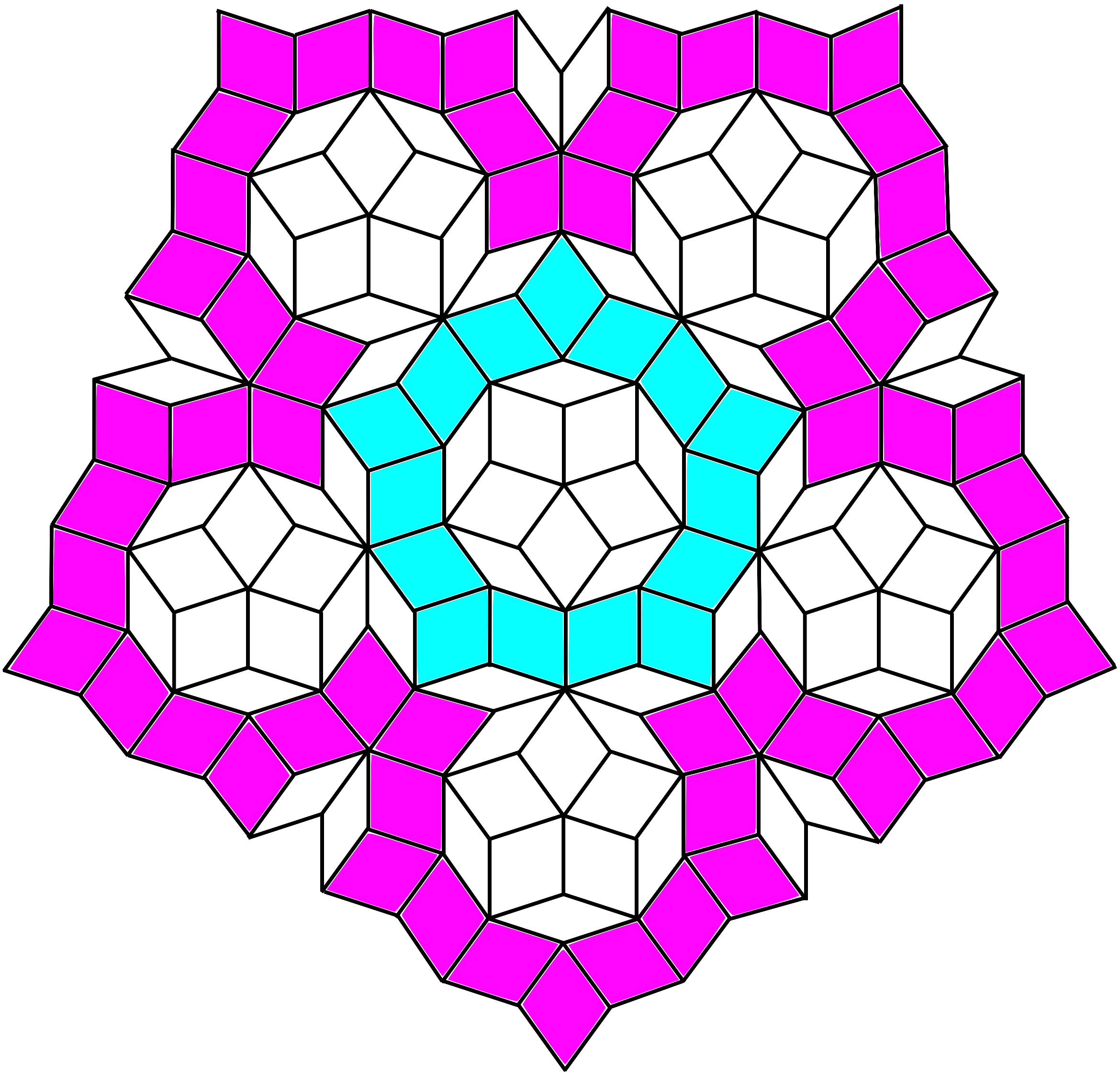}   
\end{center}
\vspace{3mm}
\begin{center}
\includegraphics[width=\linewidth]{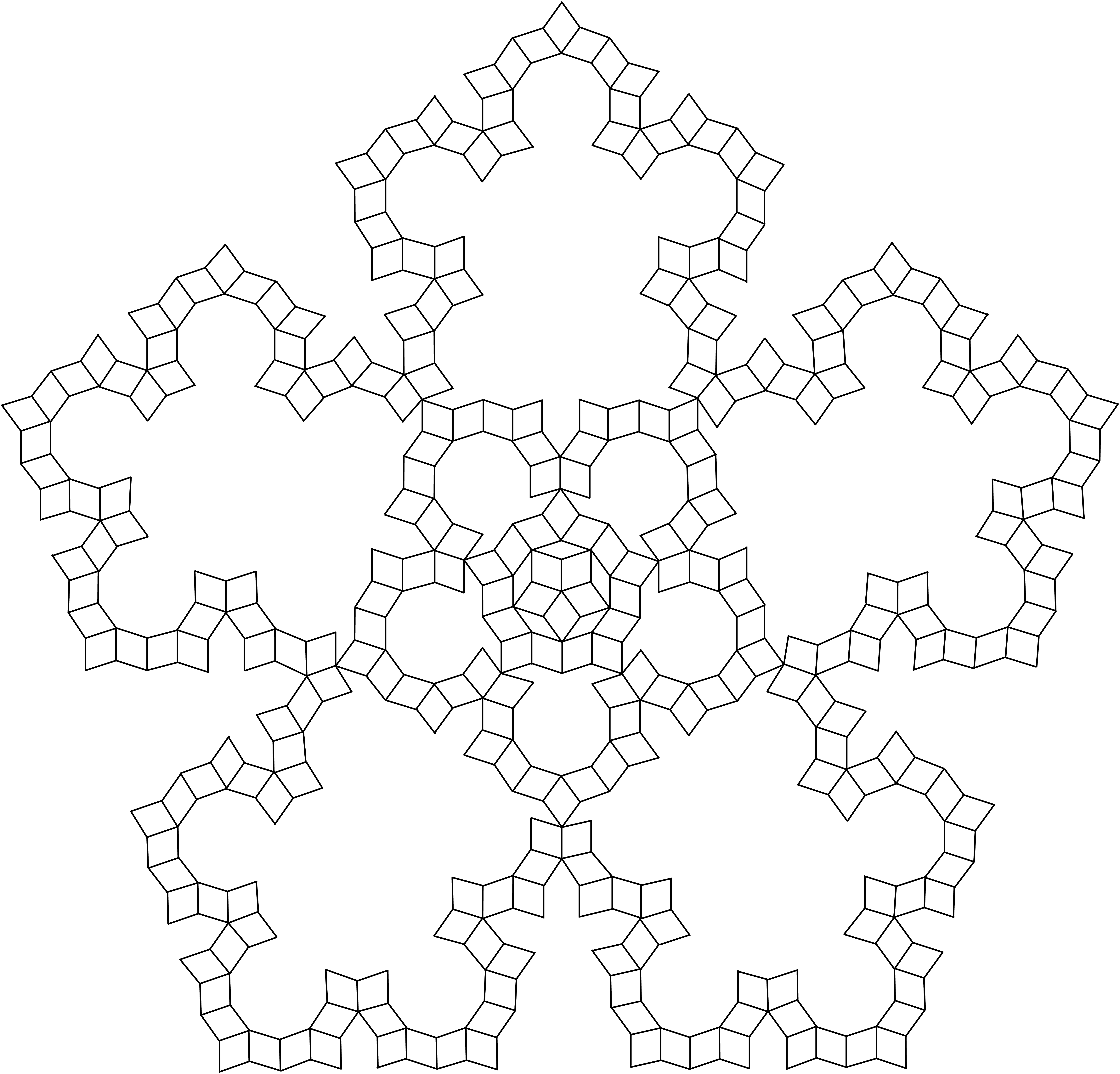} 
\end{center}
\caption{Concentric circles formed by ``thick'' rhombi in the Penrose 
tiling.}
\label{PenroseColor}
\end{figure}

\begin{figure}[t]
\begin{center}
\includegraphics[width=\linewidth]{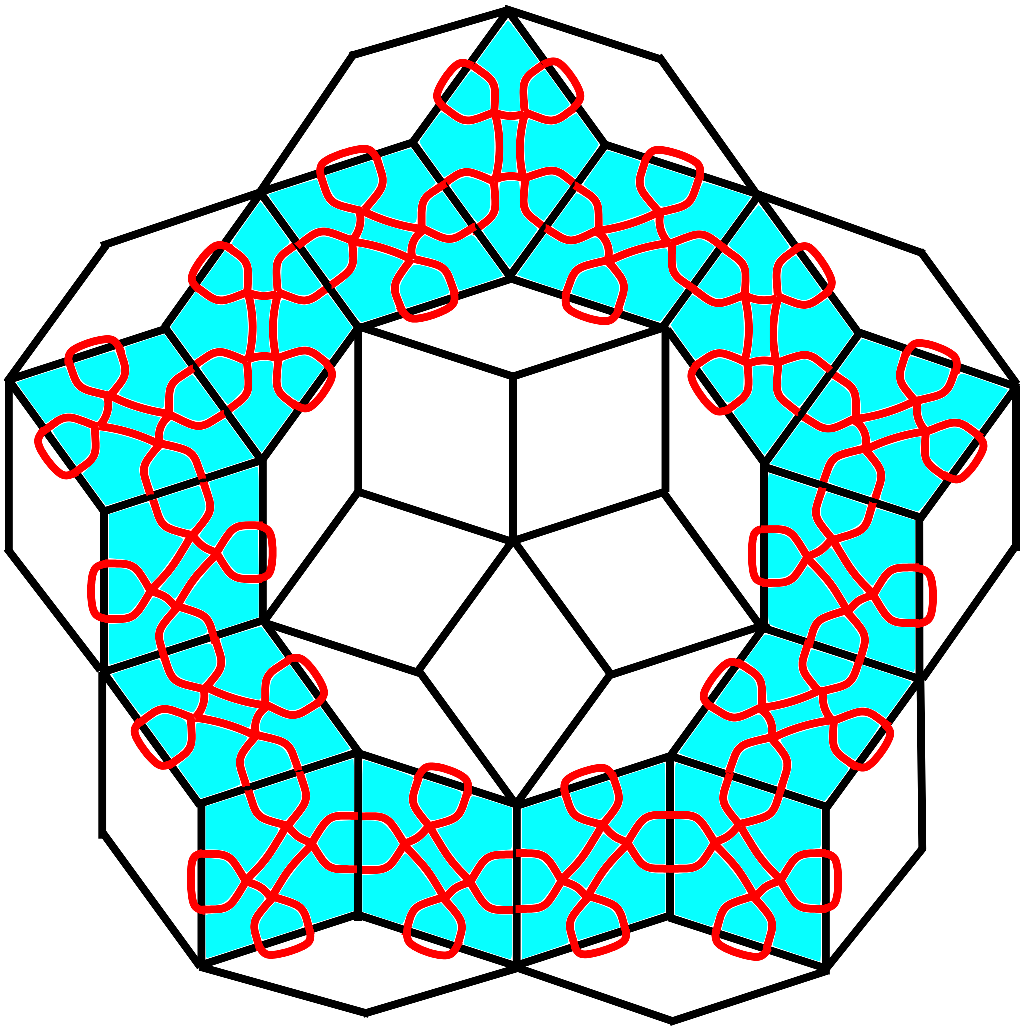}
\end{center}
\caption{Singular level lines $\, f(x, y) = f_{1} \, $ of large diameter.}
\label{PenroseNet}
\end{figure}

\begin{figure}[t]
\begin{center}
\includegraphics[width=\linewidth]{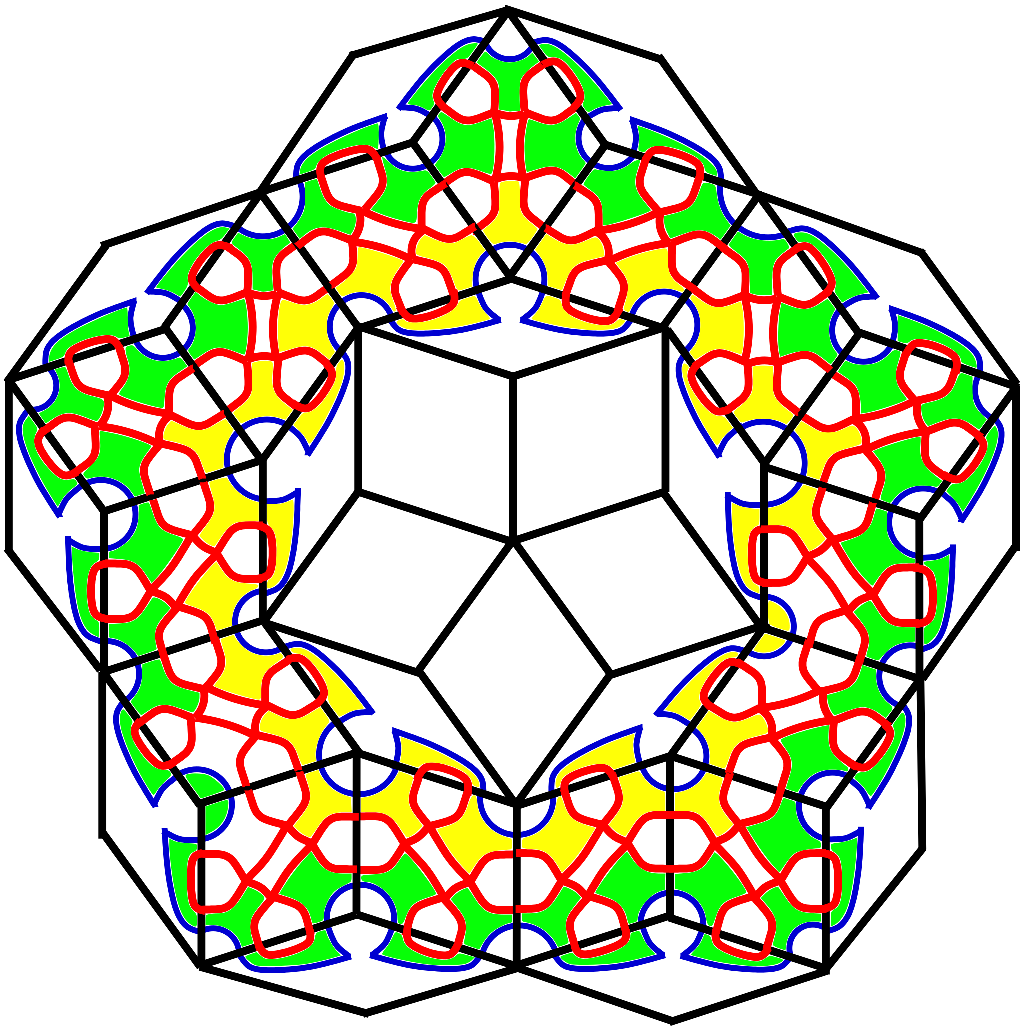}
\end{center}
\caption{Layers of non-singular level lines of $\, f (x, y) \, $ in 
the range $\, f_{1} < f (x, y) < f_{2} \, $.}
\label{PenroseLines}
\end{figure}

 Coming back to the tiling with exact rotational symmetry, one can observe the 
existence of concentric rings of increasing diameter, formed by ``thick'' rhombi 
(Fig. \ref{PenroseColor}). It can also be seen that each such ring corresponds 
to an annular singular conglomerate at the level $\, f (x, y) = f_{1} \, $
(Fig. \ref{PenroseNet}). This conglomerate is surrounded by layers of non-singular 
level curves $\, f (x, y) = c > f_{1} \, $ of the same size, which exist within 
the energy interval $\, c \in \left( f_{1} , f_{2} \right) \, $. At the value 
$\, c = f_{2} \, $ these layers connect to other conglomerates associated with 
the second set of saddle-type singular points of the function $\, f (x, y) \, $ 
(Fig. \ref{PenroseLines}).

 It can then be observed that the presented Penrose tiling can be 
associated with a function $\, f (x, y) \, $ that has no open level lines; 
however, for $\, c \in \left[ f_{1} , f_{2} \right] \, $ it does possess 
closed level components $\, f (x, y) = c \, $ of arbitrarily large diameter. 
As follows from the results of the previous chapter, among other tilings 
(obtained via parallel shifts of $\, \mathbb{R}^{2} \, $ within 
$\, \mathbb{R}^{4}$), for any value $\, c \in \left[ f_{1} , f_{2} \right] \, $,
there must exist tilings that contain open level lines $\, f (x, y) = c \, $.
These open level lines of $\, f (x, y) \, $ are non-singular within the 
interval $\, c \in \left( f_{1} , f_{2} \right) \, $, but they transform 
into singular conglomerates at $\, c = f_{1} \, $ and $\, c = f_{2} \, $.
For values $\, c < f_{1} \, $ or $\, c > f_{2} \, $, these singular 
conglomerates break apart into closed level components of the function 
$\, f (x, y) \, $. 

 As we mentioned above, such a situation is impossible for A-type 
quasiperiodic functions in $\, \mathbb{R}^{2} \, $ that possess dihedral 
(quasicrystalline) symmetry (\cite{Sirius}).

\vspace{2mm}

 The example above illustrates also another distinction in Novikov’s problem 
between quasiperiodic functions of type A and type B. Namely, for both type A 
and type B functions, the emergence of open level lines (\ref{LevelSurf}) in 
at least one plane $\, \mathbb{R}^{2} \subset \mathbb{R}^{N} \, $ of a given 
direction corresponds to a closed interval 
$\, c \in \left[ c_{1}, \, c_{2} \right] \, $. 
However, for type A functions, the absence of open level lines in certain 
planes $\, \mathbb{R}^{2} \subset \mathbb{R}^{N} \, $ of this direction - even 
in the presence of  bounded level lines of arbitrarily large size - can only occur 
at the endpoints $\, c = c_{1} \, $ or $\, c = c_{2} \, $ \cite{BigQuas}. 
In the example above, there exists a plane 
$\, \mathbb{R}^{2} \subset \mathbb{R}^{4} \, $ that contains only bounded level 
lines of arbitrarily large size for all values 
$\, c \in \left[ f_{1}, \, f_{2} \right] \, $.

\vspace{2mm}

 In conclusion, we would also like to note that it would be quite 
interesting to construct examples of functions in $\, \mathbb{R}^{2} \, $ 
with 3 quasi-periods (of type B) that possess chaotic (i.e., not 
topologically regular) open level lines (\ref{LevelSurf}) within 
a certain finite interval of values $\, c \, $.

\section{Conclusion}
\setcounter{equation}{0}

 The paper examines two distinct classes of smooth quasiperiodic 
functions in $\, \mathbb{R}^{n} \, $ that arise in modern theoretical 
and mathematical physics. Functions of the first type (type A) emerge 
as restrictions of smooth periodic functions in higher-dimensional spaces 
to a subspace $\, \mathbb{R}^{n} \, $ via an affine embedding 
$\, \mathbb{R}^{n} \subset \mathbb{R}^{N} \, $. Functions of the second 
type (type B) arise as potentials and fields generated by quasicrystalline 
structures in $\, \mathbb{R}^{n} \, $. Both types of functions play 
a crucial role in a variety of physical systems and appear across a wide 
range of contemporary theoretical and applied problems. This work 
investigates the problem posed by S.P. Novikov - namely, the 
characterization of the level sets of quasiperiodic functions - for 
both types. As demonstrated in the paper, potentials of type A and type B 
share several common properties, yet they also exhibit a number of 
significant differences in the context of this problem.

\end{document}